\documentclass[12pt,a4paper]{article}

  \usepackage{a4wide}
  \usepackage{latexsym}
  \usepackage{epsf}
  \usepackage{amssymb}
  \usepackage{graphicx}
  \usepackage{amsmath, cite}
  \usepackage{amsmath,amssymb,amsthm}
  \usepackage{verbatim}
  \usepackage{hyperref}

\renewcommand{\d}{\textrm{d}}
\newcommand{\e}{\textrm{e}}

\newcommand{\w}{\wedge}

\newcommand{\SO}{\mathop{\rm SO}}
\newcommand{\SU}{\mathop{\rm SU}}
\newcommand{\U}{\mathop{\rm U}}

\newcommand{\kper}{k_{\perp}}
\newcommand{\kpar}{k_{\parallel}}
\renewcommand{\Re}{\operatorname{Re}}
\renewcommand{\Im}{\operatorname{Im}}

\newcommand{\be}{\begin{equation}}
\newcommand{\ee}{\end{equation}}
\newcommand{\beq}{\begin{equation}}
\newcommand{\eeq}{\end{equation}}
\newcommand{\ba}{\begin{eqnarray}}
\newcommand{\ea}{\end{eqnarray}}
\newcommand{\bea}{\begin{eqnarray}}
\newcommand{\eea}{\end{eqnarray}}
\newcommand{\nn}{\nonumber}

\renewcommand{\d}{\textrm{d}}

\begin{document}
\numberwithin{equation}{section}

\begin{center}

{\LARGE \bf{AdS vacua with scale separation \\ \vspace{0.5cm} from IIB supergravity}}

\vspace{1.1 cm} {\large M. Petrini$^a$, G. Solard$^a$ and T. Van
Riet$^b$}\\

\vspace{0.8 cm}{$^{a }$ Laboratoire de Physique Th\'eorique et Hautes Energies,\\
Universit\'e Pierre et Marie Curie,\\ 4 Place Jussieu, 75252 Paris Cedex 05, France}\\
\vspace{.1 cm} {$^b$ Instituut voor Theoretische Fysica, K.U. Leuven,\\
Celestijnenlaan 200D B-3001 Leuven, Belgium} \footnote{{\ttfamily {petrini@lpthe.jussieu.fr, solard@lpthe.jussieu.fr,  thomasvr @ itf.fys.kuleuven.be}}}\\

\vspace{0.8cm}

{\bf Abstract}
\end{center}

\begin{quotation}
Only two kinds of compactification are known that lead to four-dimensional supersymmetric AdS vacua with moduli stabilisation and separation of scales at tree-level. The most studied ones are compactifications of massive IIA supergravity on $\SU(3)$ structures with smeared O6 planes, for which a general ten-dimensional expression for the solution in terms of the $\SU(3)$ structure was found. Less studied are compactifications of IIB supergravity with smeared O5/O7 planes. In this paper we derive a general ten-dimensional expression for the smeared O5/O7 solutions in terms  of $\SU(2)$ structures.
For a specific choice of orientifold projections, we recover the known  examples and we also provide new explicit solutions.

\end{quotation}
\newpage

\tableofcontents

\section{Introduction}
One of the ultimate goals of the research  on flux compactifications is the construction of flux vacua that break supersymmetry in a solution with a small positive cosmological constant. Despite many interesting ideas and proposals, it is still debatable whether there exist fully explicit and controllable solutions to this problem. A possible strategy is to break up the problem in parts and solve each  part separately. In this paper we therefore settle with studying SUSY AdS vacua and set aside the issue of SUSY breaking and positive cosmological constant.   Despite the fact that SUSY AdS vacua are the best known and most constructed solutions in string theory, there are still some obvious and interesting problems as we point out below.

\vspace{0.3cm}

Before the original KKLT proposal \cite{Kachru:2003aw} (and \cite{Balasubramanian:2005zx}), none of the AdS vacua in string theory were truly lower-dimensional in the sense that the AdS scale was not  parametrically larger than the length scale of the extra dimensions. This is the most straightforward constraint that observations put on flux vacua.  The AdS solutions that are used for holography typically  do not have scale separation and it is important to understand how holography works for AdS vacua with scale separation \cite{Polchinski:2009ch}. The KKLT construction and its descendants are not entirely explicit from a 10-dimensional point of view, which complicates a possible holographic understanding. For that reason, and for reasons of elegance and simplicity, it would be desirable to have solutions of classical supergravity in ten dimensions.
This was first claimed in a series of papers constructing  such vacua in massive IIA supergravity with intersecting O6 planes \cite{Derendinger:2004jn,  Behrndt:2004km, DeWolfe:2005uu} (see also \cite{Lust:2004ig, Caviezel:2008ik} for later work on these solutions). In massive IIA many solutions without sources are also  known \cite{Tomasiello:2007eq, Koerber:2010rn}, but they cannot  achieve scale separation \cite{Tsimpis:2012tu}. Only for solutions with O6 planes is this possible, although no no-go theorem 
excluding other possibilities  has been found. 

It is unfortunate that the orientifold and D-brane sources in these compactifications are not fully understood. Most prominent is the fact that in most cases the supergravity solutions are constructed in the limit where  the sources are  smeared over the transverse space (see for instance the discussion in \cite{McOrist:2012yc}). The way the solution is supposed to change for  fully localised sources is still an open question, even if some interesting progress was made in \cite{Saracco:2012wc}.  For compactifications with sources that are \emph{parallel} (or have an F-theory interpretation), such as for the no-scale orientifold compactifications of \cite{Dasgupta:1999ss, Giddings:2001yu} and their T-duals \cite{Grana:2006kf, Blaback:2010sj, Blaback:2012mu },  it is known how to treat fully localised sources. For these cases the backreaction does not invalidate the existence of the solutions, but it is expected to be very relevant when computing fluctuations around the vacuum (see for instance \cite{Shiu:2008ry, Frey:2008xw, Frey:2013bha}).
However the AdS compactifications in IIA with scale separation involve \emph{intersecting} O6 planes and, apart from the partial results in \cite{Saracco:2012wc},  not much is known.  Another issue that troubles these vacua  is more stringy and concerns the proper definition of string theory with O6 planes when there is non-zero Romans mass \cite{Banks:2006hg, McOrist:2012yc}. Since there is no conventional lift of massive IIA supergravity to 11 dimensions\footnote{See however the intriguing proposal of \cite{Bergshoeff:1997ak}, or the alternative suggestion that a lift is unnecessary since massive IIA cannot be strongly coupled at weak curvature \cite{Aharony:2010af}. } it is not clear how the orientifold singularities can be resolved and whether the background makes sense\footnote{See reference \cite{Copsey:2013jla} for more radical doubts about the use of orientifold planes.}. 

\vspace{0.3cm}

For the reasons just named it is relevant to find other classical AdS solutions with scale separation in a different context. In this paper we consider type IIB supergravity (string theory). It is commonly claimed that this theory cannot achieve moduli stabilisation at the classical level, but this statement can readily be violated by considering non-geometric fluxes or by moving beyond the usual O3/O7 compactifications and instead relying on O5/O7 orientifolds.  A first attempt at finding such vacua has been done
in \cite{Caviezel:2009tu}, where the authors considered four dimensional effective theories obtained by consistent truncations on 
specific $SU(2)$ structure manifolds  (built from coset space coverings) with smeared O5/O7 intersections.   While some of the models considered allow for full moduli
stabilisation, it is not clear whether they admit a limit in which the solution is at large volume, weak coupling and with scale separation.

The aim of this paper is to further study O5/O7 compactifications of IIB supergravity to four-dimensional, unwarped  AdS space, the absence of warping being a necessary outcome of the approximation of smeared sources.  Our results have partial overlap with an earlier investigation on SUSY AdS vacua in IIA/IIB SUGRA \cite{Lust:2009zb}.  
We construct the solutions directly in ten dimensions  using the pure spinors approach proposed in \cite{Grana:2004bg, Grana:2005sn, Grana:2006kf}.

 For compactifications to four dimensions, this formalism allows to reduce the study of ten-dimensional supersymmetric backgrounds to the analysis of a set of equations involving only the components of the fields on the internal manifold.  In this case, it is easy to show that the O-plane projections and supersymmetry require the internal manifold to admit a rigid  $SU(2)$
structure. It is then possible to  write down a general solution for  the fields on the compactification manifold.  
By general solution we mean a set of constraints 
on the  six-dimensional fields that are applicable to a whole class of manifolds instead of a specific example. This is typically achieved through writing the solution in terms of the 
$SU(2)$ invariant  forms on the manifolds.  To go from this general  form to a concrete example one only has to compute the canonical forms for a given manifold. This is clearly beneficial and more insightful than minimizing $F$ and $D$ terms for a given manifold. When the compactification manifolds allow 
for consistent truncations, which is the case for homogeneous manifolds with smeared sources, then the minima of the scalar potential must lift consistently to solutions of the equations of motion in ten dimensions, such as derived in this paper.  Reference \cite{Acharya:2006ne} explicitly analysed how the IIA vacua in 4D lift to 10 IIA SUGRA solutions with smeared sources. 

Our analysis parallels the derivation given in  \cite{Grana:2006kf} of the conditions for $\mathcal{N}=1$  AdS$_4$ vacua 
of \cite{Lust:2004ig}. However the IIB case seems to have a much richer spectrum of solutions than the IIA case.
For this reason,   in looking for explicit examples, we will restrict to homogeneous spaces.
We perform a systematic scan of the  coset manifolds  of \cite{Koerber:2008rx}  and nilmanifolds.
Although we have not been able to find solutions fulfilling the criteria of weak coupling and scale separation on the cosets, we have found such solutions on nilmanifolds thereby extending the single solution that was known so far \cite{Caviezel:2008ik}, which was obtained by T-dualising the O6 solution on the six-torus.   Although these new solutions are also incomplete in the sense that the sources are  smeared, they constitute an important step  since they change the existing paradigm that tree-level scale separation is only possible in massive IIA. Besides providing more examples of tree-level scale separation, the examples in IIB could be relevant for improving our understanding of the subtleties in such backgrounds, such as the orientifold singularities. The usual criticism of using O6 planes in the presence of Romans mass is clearly evaded here.


\vspace{0.3cm}

The paper is organised as follows. In Section \ref{gencon} we give the general conditions that the fields on the internal manifold have to satisfy in order
to have $\mathcal{N}=1$  AdS$_4$ with O5/O7 planes.  Since the derivation is quite lengthy we put it in Appendix \ref{SUSYcondps}.  In Section \ref{simplesol}
we specify the general system to a class of $SU(2)$ structures that are for instance allowed on homogeneous spaces. The orientifold projections and the restriction to
left-invariant forms make it possible to solve explicitly for most of the constraints. In Section \ref{scalesep} we discuss possible criteria to check whether a given
vacuum admit separation of scales. Finally in Section \ref{examples} we present an exhaustive list of $\mathcal{N}=1$  AdS$_4$ with O5/O7 planes that can
be found on cosets and nilmanifolds and we discuss their properties as good 4-dimensional vacua.  

Appendix \ref{SUSYcondps} contains the definitions of  $\SU(3)$ and $\SU(2)$ structures and pure spinors we need in the rest of the paper, and
the derivation of the general conditions for $\mathcal{N}=1$ AdS$_4$ susy vacua.  For completeness in Appendix \ref{IIAapp} we give the form
of the generic $\mathcal{N}=1$ AdS$_4$ vacuum in type IIA and discuss an example of separation of scale in this context.
Finally, in  Appendix \ref{riccisapp} we detail de form of the Ricci scalar for the class of $\SU(2)$ solutions we consider in this paper.

\section{Type IIB AdS$_4$ vacua with $\mathcal{N}=1$ supersymmetry}
\label{gencon}

A standard technique to study supersymmetric vacua is to look for solutions to the supersymmetry variations plus the Bianchi identities for the fluxes.
In presence of non-trivial backgrounds fluxes, instead of working directly with spinorial equations, it is more 
convenient to rewrite the susy equations as a  set of 
differential conditions on forms.  This is the idea behind the application of $G$-structures and more generally  Generalised Complex Geometry  \cite{Hitchin:2004ut, Gualtieri:2003dx}. We begin this section with a brief overview of the formalism we need to determine our solutions. Details can be found in Appendix \ref{SUSYcondps}.

\vspace{0.3cm}

In the Generalised Complex Geometry approach the main ingredients are a pair of polyforms, $\Phi_\pm$, which are constructed as bilinears in the
supersymmetry parameters on the internal manifold $Y$
\beq
\label{puresp}
\Phi_\pm  = \eta^1_+ \otimes \eta^{2 \dagger}_\pm\, ,
\eeq
where $\eta^i_+$ are six-dimensional Weyl spinors, and $\eta^i_- = (\eta^i_+)^\ast$.  
Then, the ten-dimensional supersymmetry variations can be rewritten as a set of differential conditions on such forms. For Type IIB compactifications to AdS$_4$, the
susy conditions are \cite{Grana:2004bg}
\bea
\label{eqm}
&&(\d - H  \w )(e^{2A - \phi} \Phi_-)  = - 2 \mu e^{A-\phi} {\rm Re } \Phi_+,  \\
\label{eqpa}
&&  (\d - H \w )(e^{A -\phi } {\rm Re} \, \Phi_+) =0 \, , \\
\label{eqpb}
&& (\d - H \w )(e^{3 A -\phi } {\rm Im} \, \Phi_+) = -3e^{2 A-\phi}{\rm Im} \, (\bar{\mu}\Phi_-) -\frac{1}{8} e^{4 A} \ast \lambda(F) \,, 
\eea
 where $\phi$ is the dilaton, $A$ the warp factor 
 \beq
 \label{10dmetric}
 \d s^2 = e^{2 A} \d s^2_{(4)} + \d s^2_{(6)} \, ,
 \eeq
and $F$  is the sum of the RR field strength on  $Y$,  $F = F_1 + F_3 + F_5$.  $\lambda$ acts on a form as  the transposition of all indices
 \beq
 \label{lambdaop}
 \lambda(F_k) = (-)^{[k/2]} F_k \, .
 \eeq
The ten-dimensional fluxes are defined in terms of $F$ by
\beq
F_{(10)} = {\rm vol}_4 \wedge \lambda (\ast F) + F \, .
\eeq
Finally, the complex number $\mu$ determines the size of the AdS$_4$ cosmological constant 
\beq
\Lambda=-|\mu|^2\,.
\eeq

\vspace{0.3cm}

The form of the pure spinors $\Phi_\pm$ depends on the relation between the internal spinors
$\eta_1$ and $\eta_2$. In the most general case we have
\bea
\label{Phim}
&& \Phi_-= -\frac{ab}{8}   z \wedge (\kper e^{ -ij }  + i \kpar \omega ) \, , \\
\label{Phip}
&& \Phi_+=  \frac{ a \bar{b}}{8}   \, e^{ z \bar{z} /2} (\kpar e^{-ij}  -i\kper \omega ) \, ,
\eea
where $z$, $j$ and $\omega$ are a one-form, a real two-form and  a holomorphic two-form, respectively, which are globally defined on the internal manifold
and define a $\SU(2)$ structure\footnote{The  definition of $\SU(2)$ and $\SU(3)$ structures can be found in Appendix \ref{SUSYcondps}.}.  
The complex functions $a$ and $b$ are related to the norms of the spinors $\eta^i$
\bea
&& \| \eta^1_+ \|^2 = |a|^2\,, \qquad \qquad  \| \eta^2_+\|^2= |b|^2 \, ,
\eea
and to the norm of the pure spinors 
\beq
\label{normps}
\langle \Phi_\pm , \bar{\Phi}_\pm \rangle = - i \| \Phi_\pm \|^2 {\rm vol}_6 =  - \frac{i}{8}  | a |^2  |b|^2 {\rm vol}_6 \, ,
\eeq
where ${\rm vol}_6$ is the volume of the internal manifold and the product 
\beq
\label{Miukai}
\langle A , B  \rangle  = (A \wedge \lambda(B))|_{\rm top}
\eeq
is the Mukai pairing among forms.

When $\kpar = 1$ and $\kper=0$  the spinors $\eta_+^1$ and $\eta_+^2$  are parallel and the internal manifold is said to be of $\SU(3)$ structure.
The pure spinors reduce to
\bea
\label{SU3pstext}
&& \Phi_-= - i \frac{a b}{8}  \Omega  \, , \\
&& \Phi_+=  \frac{ a \bar{b}}{8} \,   e^{-iJ}   \, ,
\eea
where $\Omega$ and $J$ are the $\SU(3)$ invariant forms defining the $\SU(3)$ structure, \eqref{SU3strf}.  In the opposite case,
$\kpar = 0$ and $\kper=1$,  the spinors $\eta_+^1$ and $\eta_+^2$ are orthogonal and the structure is $\SU(2)$ with
\bea
\label{SU2pstext}
&& \Phi_-= -\frac{a b}{8} z \wedge e^{ -ij }  \, , \\
&& \Phi_+= -i  \frac{ a \bar{b}}{8} \, e^{ z \bar{z} /2}  \omega  \, .
\eea
The general case,  where the relative orientation of $\eta_+^1$ an $\eta_+^2$ can vary on the manifold is often referred to as dynamical $\SU(2)$ structure. 

\vspace{0.3cm}

For AdS vacua, supersymmetry constraints the norms of the two six-dimensional spinors to be equal \cite{Grana:2006kf}
\bea
 |a|^2 = |b|^2 = e^A \, .  
\eea
Only the relative scale between the spinor being relevant, we can always rescale $\eta_+$ in such a way that
\beq
\bar{b} =  a \,,   \qquad \qquad  \frac{b}{a}=  e^{- i\theta} \, . 
 \eeq

\vspace{0.3cm}

Equation  \eqref{eqpb} can be seen as a definition of the RR fluxes that are compatible with $\mathcal{N}=1$ supersymmetry.  In order to have a full solution of the ten-dimensional 
equations of motion, one must also check that the RR fluxes determined this way satisfy the Bianchi identities
\bea
\label{BINS}
&& \d H = 0 \, , \\
\label{BIRR}
&& \d F - H \wedge F =  \delta(sources) \, ,
\eea
where $\delta(sources)$ denotes the charge density of the space-filling sources. In this paper we will mostly consider space-filling O5 and O7-planes intersecting on
the internal manifold.  Since we do not know how to find exact solutions that describe generic intersecting branes or O-planes, we smear them over the internal manifold, and we write the source terms as invariant smooth forms on the internal manifold\footnote{We refer to Appendix C of \cite{Caviezel:2008ik} for an explanation on smeared source terms and the corresponding microscopic interpretation in terms of orientifolds and their involutions, whereas Appendix D of \cite{Danielsson:2011au} contains some first attempts for charge and flux quantisation. }
\beq
\label{BIRRsmeareds}
 \d F - H \wedge F =  \sum_i c_i \alpha^i = \sum_i Q_i(source) {\rm vol}_i
\eeq
where $c_i$  are constants and $Q_i$  is the charge density of the source. The symbol  $\alpha^i$ denote a decomposable form dual to the cycle wrapped by the brane, while
${\rm vol}_i$ is the volume form dual to the cycle.

\subsection{General constraints}

Plugging the expression \eqref{SU3pstext} of the pure spinors  in the SUSY equations, it is easy to see that,  for $\SU(3)$ structure manifolds,  \eqref{eqm} has no solutions. 
We recover the known result that  there  are no $\mathcal{N}=1$ AdS$_4$ vacua with $\SU(3)$ structure in type IIB supergravity \cite{Behrndt:2005bv}.

\vspace{0.3cm}

We are left with the possibility of rigid or dynamical $\SU(2)$ structure. By expanding  the supersymmetry equations  in forms of definite degree we can package the conditions for AdS$_4$ vacua with $\mathcal{N}=1$  supersymmetry as a set of differential constraints  on the $\SU(2)$ structure forms, the fluxes and the functions $\kpar$ and $\kper$. 
Let us consider again  \eqref{eqm}.  The zero-form  component 
\beq
\mu \,  \kpar \cos \theta = 0 \, ,
\eeq
gives a constraint on the parameters $\kpar$ and $\theta_-$, since for AdS vacua $\mu \neq  0$. Choosing
\beq
\cos \theta  = 0 \, ,
\eeq
fixes the relative phase of the spinors $\eta^1$ and $\eta^2$
\beq
\theta  =  \theta_a -  \theta_b  =  \frac{\pi}{2} \qquad  \Rightarrow \qquad  a = i b \, .
\eeq 
Such phase  is related to the choice of orientifold planes that one can add as sources, see for instance \cite{Grana:2006kf, Koerber:2007jb}. On can show that
 $a = i b$ is compatible with O7-planes only. Since we want to be free to have also O5-planes, we are forced to choose the first option and set
\beq
\kpar = 0 \qquad \kper =1 \, ,
\eeq
which means that we will only consider backgrounds of rigid $\SU(2)$-structure. For this case the supersymmetry conditions reduce to a set of
equations for the $\SU(2)$ structure forms
(see again Appendix
\ref{SUSYcondps} for the derivation)
\bea
\label{Phimexpf1tx}
&& \d (e^{3 A - \phi} z) =  2 \mu e^{2 A - \phi}  {\rm Im} \, \hat{\omega} \,  , \\ 
\label{Phimexpf2tx}
&& z \w ( \d j - i H + \mu e^{-A} \, \bar{z} {\rm Re} \, \hat{\omega}  ) =0  \\
\label{Phipaexpf1tx}
&& \d (e^{2 A - \phi}  {\rm Im} \hat{\omega}  ) = 0 \, , \\
\label{Phipaexpf2tx}
&&  \d( e^{2 A - \phi}  z \w \bar{z}  \w  {\rm Re} \hat{\omega}  ) = 2 i  e^{2 A - \phi}   H \w  {\rm Im} \hat{\omega}   \, ,
\eea
and the following equations for the RR fluxes 
\bea
\label{5formfl}
&& \ast F_5 = - 3 e^{-A - \phi} \, {\rm Im } (\bar{\mu} z) \, , \\
\label{3formfl}
&& \ast F_3 =- e ^{ -4 A} \, \d (e^{4 A - \phi} {\rm Re} \hat{\omega} ) -3 e^{-A - \phi} {\rm Re} (\bar{\mu} z) \w j  \, , \\
\label{1formfl}
&&  \ast F_1 = i \, \d (2 A - \phi)  z \w \bar{z}  \w {\rm Im} \hat{\omega}  +   e^{-\phi} H \w {\rm Re} \hat{\omega} \nn \\
& &  \qquad  \quad -
\frac{1}{2} e^{-A - \phi} {\rm Im} (\bar{\mu} z ) \w j \w j \,.
\eea
To  simplify the notation, we defined the form  $\hat{\omega} =  e^{i \theta} \omega$.  As mentioned before, in order to find solutions, we have to add to this set of constraints the Bianchi identities for the fluxes.

\subsection{Supersymmetry and $\SU(2)$ torsion classes}

To make contact with previous literature,  we can express the equations above in terms of  $\SU(2)$ torsion classes. The idea here is 
 to decompose all fields in the supersymmetry variations into irreducible representations of the $\SU(2)$ structure group.  
In general the forms $z$, $j$ and $\omega$ are not closed. Their deviation from closure can be expressed in terms of different $\SU(2)$ representations, called the torsion
classes\footnote{See \cite{Gauntlett:2002sc, Gauntlett:2003cy} for a detailed discussion of intrinsic torsion.}  \cite{Dall'Agata:2003ir}
\bea
\label{SU2torsionstxt}
&& \d z = S_1 \omega +S_2j+S_3 z \wedge \bar{z}+S_4\bar{\omega} + z \wedge (V_1+\bar{V}_2)+\bar{z}  \wedge (V_3+\bar{V}_4)+T_1 \nn \\
&& \d j= S_5 \bar{z} \wedge \omega +S_6  z \wedge \omega + \frac{1}{2}(S_7 + \bar{S}_8)  z \wedge j +j \wedge V_5 + z \wedge \bar{z} \wedge V_6
+ z  \wedge T_2+\mathrm{c.c.} \nn \\
&& \d \omega = S_7 z \wedge \omega + S_8 \bar{z} \wedge \omega -2 \bar{S}_{5} z \wedge j -2 
\bar{S}_{6} \bar{z} \wedge j  + i z \wedge \bar{z} \wedge (\bar{V}_6  \llcorner \omega) +j \wedge (V_7+\bar{V}_{8})  \nn \\ 
&& \qquad +  z \wedge T_3 +\bar{z} \wedge T_4  \, .
\eea
The coefficients $S_i$,  $V_i$ and $T_i$ denote the 20 different $\SU(2)$ torsion classes:  8 complex singlet $S_i$, 8 complex doublets\footnote{We added two vector representations  in $\d z$ that were missing in \cite{Dall'Agata:2003ir}.}  $V_i$ and
4 complex triplets $T_i$.  

\vspace{0.3cm}

The NS and RR fluxes can also be decomposed according to $\SU(2)$ representations
\begin{align}
 H= &h_1 z \wedge \hat{\omega}+h_2\bar{z} \wedge \hat{\omega}
+h_3 z  \wedge j+ z \wedge \bar{z} \wedge h_1^{(2)}+h_{2}^{(2)}\wedge j + z \wedge  h^{(3)}+\mathrm{c.c.}\\
F_1=&f_1 z+f_1^{(2)}+\mathrm{c.c.}\\
F_3=&f_2 z \ \wedge \hat{\omega}+f_3\bar{z} \wedge  \hat{\omega}+f_4 z \wedge j+ z \wedge \bar{z} \wedge f_2^{(2)}+f_3^{(2)} \wedge j+ z \wedge f^{(3)}+\mathrm{c.c.}\\
F_5=&f_5 z \wedge j \wedge j + z \wedge \bar{z}  \wedge j \wedge f_4^{(2)}+\mathrm{c.c.}
\end{align}
where  $h_i$ and $f_i$ are complex scalars in the singlet representation of $\SU(2)$,  $h_i^{(2)}$ and $f_i^{(2)}$ are holomorphic vectors in  the ${\bf 2}$ and $h^{(3)}$ and
$f^{(3)}$ are complex two forms in the triplet representation, which are (1,1) and primitive with respect to $j$. 

\vspace{0.3cm}

Using the above decompositions, the supersymmetry variations can be written as a set of conditions on the torsions classes and the fluxes.
The singlets in the  torsions must satisfy 
\beq
\label{singlets}
\begin{array}{lll}
S_2= 0 \, , &  \qquad  S_1 = - S_4 = -i \mu e^{-A}  \, , \\
S_3=\frac{1}{2}\partial_{\bar{z}}(3A-\phi) \, ,    & \qquad   S_5 = \bar{S}_6 = i \bar{h}_1 -\frac{1}{2} e^{-A}\mu  \, , \\
& \qquad S_7=\bar{S_8} = - \frac{1}{2} \partial_{z}(2A-\phi)   \, , & 
\end{array}
\eeq
while  the vectors  are 
\beq
\label{vectors}
\begin{array}{ll}
V_3=V_4= V_6 = 0  \, ,   & \quad   V_7  = i  (\bar{\partial}_4 A  + \bar{h}^{(2)}_1 )  \llcorner  \omega  \, , \\  
V_5=i h_2^{(2)}  \, ,  &  \quad V_8 = i     [ \bar{\partial}_4 (3 A -\phi)  + \bar{h}^{(2)}_1 ]  \llcorner  \omega  \, , \\
 V_1=V_2=\partial_4 (3A-\phi)  \, ,  & \qquad  
\end{array}
\eeq
and the two-forms read
\beq
\label{2forms}
T_1 = 0  \, ,  \qquad T_2 = - i h^{(3)} \, , \qquad T_3 = \bar{T}_4  \, .
\eeq

For the fluxes, we find for  the  NS flux singlets 
\beq
h_1 =\bar{h}_2 \, , \qquad h_3 =  -  \frac{i}{2} \partial_z (2 A - \phi) \, ,
\eeq
and  for the RR fluxes 
\beq
\label{fluxcon}
\begin{array}{lll}
 f_1 =  e^{- \phi} (  \frac{1}{2} \bar{\mu} e^{-A} - 4 i h_1 ) \, ,  & \qquad f^{(2)}_1 = i e^{- \phi}  \omega \llcorner [\bar{\partial}_4 (2 A - \phi) + \bar{ h}^{(2)}_1 ]  \, , \\
 f_2 = \bar{f}_3 = -  \frac{i}{2} e^{- \phi} \partial_z A  \, , & \qquad   f^{(2)}_2 =  \frac{i}{2} e^{- \phi} \omega \llcorner  [\bar{\partial}_4 (4 A - \phi) - \bar{ h}^{(2)}_1 ]  \, , \\
f_4 =  \frac{1}{2}  e^{- \phi} ( 4 h_1 +  i  \bar{\mu} e^{-A} ) \, ,   & \qquad   f^{(2)}_3 = f^{(2)}_4 = 0  \, , \\
f_5 =  \frac{3}{4} e^{- A - \phi} \bar{\mu}  \, , & \qquad f^{(3)} = i  e^{-\phi} T_3  \, .
\end{array}
\eeq

 On the  fluxes given above still   we still have to impose the Bianchi identities.

\section{A simple class of $\SU(2)$ structure geometries}
\label{simplesol}

A general analysis of the $\SU(2)$ structure constraints derived in the previous section is very involved, due to the large number of torsion classes.
In this section we restrict to a  subset of all possible $\SU(2)$ structure geometries for which some of the torsion classes are set to zero.
Our motivation is to make contact with explicit examples, which are most  easily constructed on homogeneous manifolds (groups and cosets) by 
restricting to left-invariant forms.  Moreover we expect to have O-planes in our solutions as they are required to achieve
a hierarchy of scales.  The possible $\SU(2)$ structures one can define out of left-invariant forms, consistent with the orientifold involutions, is restricted. It is this restriction that we consider in this section.  Even if  we use homogeneous spaces to justify the specific choice of  $\SU(2)$ torsions, the general solutions we derive  could be applicable to more general manifolds.

\vspace{0.3cm}

Following  \cite{Caviezel:2009tu}, we introduce O5 and O7 planes filling the AdS$_4$ directions and mutually intersecting on the internal manifold.  Since they are intersecting, we take the O-planes to be smeared on the internal manifold. 
The orientifold directions are fixed by requiring  $\mathcal{N}=1$ supersymmetry \footnote{For two O-planes or D-branes to be mutually supersymmetric the number
of mixed Neumann and Dirichelet direction must be divisible by four.}  and compatibility with the  $\SU(2)$ structure.

The orientifold action on the pure spinors is given by \cite{Grana:2006kf, Koerber:2007hd}
\beq
\sigma(\Phi_+) = \pm \lambda( \bar{\Phi}_+)  \qquad \qquad \sigma(\Phi_-) = \mp \lambda(\Phi_-)
\eeq
where $\sigma$ is the orientifold involution,  $\lambda$ is the transposition operator \eqref{lambdaop} and the upper and lower signs correspond to O5 and O7 planes, respectively.
From this  we can deduce how the orientifold involution acts on the $\SU(2)$ structure forms. 
A dynamical $\SU(2)$ structure is not compatible with both O5 and O7 projections, since, when $\kpar \neq 0$ the phases of the spinors have to be different for O5 and O7 planes \cite{Koerber:2007hd}
\beq
{\rm O5} \, :  \, a = \pm b \qquad \qquad  {\rm O7} \, : \, a = \pm i b  \, .
\eeq
Therefore, all we need is the orientifold action on the rigid $\SU(2)$ structure forms
\bea
\label{Oproj}
&& \sigma (z) = \mp z  \, , \nn\\
&& \sigma(j) =  - j  \, , \nn \\
&& \sigma(\hat{\omega}) = \pm  \bar{ \hat{\omega}}   \,  .
\eea
From the equations above  we see that the one-form $z$ must be orthogonal to the O5-planes and parallel to the O7's.  It is also useful to remind how the NS and RR fluxes transform under
the orientifold involutions
\bea
\label{Oprfluxes}
&& \sigma(H) =  - H  \, , \nonumber\\
&& \sigma(F_1) = \mp F_1 \, , \nonumber \\
&& \sigma(F_3) = \pm F_3 \, , \nonumber\\
&& \sigma(F_5) = \mp F_5 \, ,
\eea
where, as before, upper and lower signs correspond to O5 and O7-planes. 

\vspace{0.3cm}

If we imagine to have a group manifold geometry, we can choose a basis of globally defined one-forms  adapted to the product structure defined by the $\SU(2)$ structure (see Appendix \ref{SUSYcondps})
and we identify the directions $e^1$ an $e^2$ with the real and imaginary part of $z$.  Then the most general choice of  O-planes is 

\begin{table}[h!]
\begin{center}
\begin{tabular}{|c||c|c||c|c|c|c|}
  \hline
  plane & 1 & 2 & 3 & 4 & 5 & 6 \\ \hline \hline
  O5 & & & x & x &  &\\ \hline
  O5 & & & & & x & x \\ \hline
  O7 & x & x & x & & x & \\ \hline
  O7 & x & x & & x & & x \\ \hline
\end{tabular}
\end{center}
\caption{O5- and O7-planes}\label{table:IIBOplanes}
\end{table}

The choice of orientifold also constrains the complex structure on $T_4 M$. One can show that the most general ansatz compatible with the orientifolds  of
Table \ref{table:IIBOplanes} is 
\bea
\label{SU2par}
&& z =z_1e^1+z_2e^2 \, , \nonumber\\
&& j=j_1e^{36}+j_2 e^{45} \,  , \nonumber \\
&& \hat{ \omega}_R= \frac{j_1j_2}{\omega_{1}}e^{34}+ \omega_{1}e^{56} \, ,\nonumber \\
&& \hat{\omega}_I=-\frac{j_1j_2}{\omega_{2}}e^{35}+\omega_{2}e^{46} \, ,
\eea
where $z_1$ and $z_2$ are complex number and $j_1$, $j_2$, $\omega_{1}$ and $\omega_{2}$ are real. 

\vspace{0.3cm}

The orientifold projections considerably simplify the $\SU(2)$ torsion classes \eqref{SU2torsions} and, consequently, the supersymmetry conditions. It is easy to see that all vector representations (doublets) in \eqref{SU2torsionstxt} are projected out.   The triplet $\tilde{j}_i$ of anti self-dual two-forms 
have the same transformation properties as $j$, $\omega_R$ and  $\omega_I$\footnote{The most general choice for the $\tilde{j}_i$ compatible with
the orientifold projection is
\beq
\begin{array}{ll}
\tilde{j}_1 = j_1e^{36} - j_2 e^{45}  \, , \\
\tilde{j}_2  =  - \frac{j_1j_2}{\omega_{1}}e^{34} + \omega_{1}e^{56}  \, , \\
\tilde{j}_3  = -\frac{j_1j_2}{\omega_{2}}e^{35} - \omega_{2}e^{46} \, . 
\end{array}
\eeq} 
\beq
\label{Optriplet}
\sigma(\tilde{j}_1) = -  \tilde{j}_1\,, \qquad  \sigma(\tilde{j}_2 ) = \pm \tilde{j}_2 \qquad \sigma(\tilde{j}_3) = \mp \tilde{j}_3 \, .
\eeq

As a result the supersymmetry conditions \eqref{singlets} - \eqref{fluxcon} reduce to 
\bea
\label{SU2torsionspar}
&& \d z = 2 \mu e^{-A} \hat{\omega}_I  \,,\nn\ \\
&& \d j=  ( 2 i \bar{h}_1 - \mu e^{- A} )  \bar{z} \wedge \hat{\omega}_R  - i  z  \wedge h^{(3)} + \mathrm{c.c.} \,,\nn \\
&& \d   \hat{\omega}_R =  ( 2 i h_1 + \bar{\mu} e^{- A} )  z \wedge j  +  z \wedge T_3 +   \mathrm{c.c.} \,,\nn \\
&& \d \hat{\omega}_I  = 0  \, ,
\eea
while the fluxes become
\bea
\label{fluxpar}
&&  H=  2 h_1 z \wedge  \hat{\omega}_R + z \wedge  h^{(3)}+\mathrm{c.c.}\,,\nn\\
&& F_1=  e^{- \phi} (4 i h_1 -\frac{1}{2}  \bar{\mu} e^{-A} )   z +\mathrm{c.c.}\,,\nn\\
&& F_3= - \frac{1}{2} e^{- \phi} (i  \bar{\mu} e^{-A} + 4  h_1)  z \wedge j + \frac{i}{2}  e^{- \phi}  z \wedge T_ 3+\mathrm{c.c.}\,,\nn\\
&& F_5=  e^{- \phi} \, f_5 z \wedge j \wedge j +\mathrm{c.c.}\,.
\eea
Comparing \eqref{Optriplet} and  \eqref{Oprfluxes} we can see that  only one of the three components survive for each $T_i$ and $h^{(3)}$
\beq
\label{partors}
T_3 = t_3 \tilde{j}_1\,, \qquad \qquad h^{(3)} = h_4 \tilde{j}_2 \, .
\eeq

In all previous equations, since we have smeared  O-planes,  we  assume that all scalars, including the warp factor and the dilaton,  are constant.

\vspace{0.3cm}

What remains to be solved are the  Bianchi identities \eqref{BINS} and \eqref{BIRRsmeareds}.  
To do so we need the derivatives of $T_3$ and $h^{(3)}$, which can be easily determined from  \eqref{partors} and
\bea
\label{dertrip}
&& \d \tilde{j}_1  =  t_3  z  \wedge \hat{\omega}_R -  a_2 z \w \tilde{j}_2 + \mathrm{ c.c}\,,\nn  \\
&& \d \tilde{j}_2 =  -i h_4   z \wedge j   + a_2  z \w \tilde{j}_1 +  \mathrm{ c.c}  \, ,
\eea
where the equations above
can obtained expanding the $\d \tilde{j}_i$ as in \eqref{SU2torsionstxt} and imposing the orientifold projections.

Let us start with the BI identities for NS three-form.  Using \eqref{SU2torsionspar}, 
 \eqref{fluxpar} and \eqref{dertrip}   we obtain 
\beq
| h_4|^2 - 4 |h_1|^2 + 2  \, {\rm Im} (e^{- A} \mu h_1    ) =0\,, \qquad \qquad  {\rm Im} ( 2 h_1 \bar{t}_3 + h_4 \bar{a}_2 ) =0  \, .\label{Finalsystem1}
\eeq

The equation for the five-form flux is trivially satisfied. We are left with the BI  involving sources (by abuse of notation we also denote by a $\delta$ the
contribution of smeared sources) 
\bea
&& \d F_1 = \delta (D7/O7)\,,\\
&&  \d F_3= H\w F_1 + \delta(D5/O5)\, .
\eea
Using again \eqref{SU2torsionspar} and \eqref{fluxpar}, they give
\bea
\label{RRBIs}
&& \delta (D7/O7) =-2  e^{ - \phi} (|e^{-A}\mu|^2  + 8 \, {\rm Im} ( e^{-A}\mu h_1) ) \hat{\omega}_I \, , \\
&&  \delta (D5/O5) = -2i e^{-\phi}(\Re(a_2\bar{t}_3 )-\Im(e^{-A}\mu h_4)-6\Re(\bar{h}_1h_4))z\wedge\bar{z}\wedge\tilde{j}_2 \nonumber\\
&& \qquad \qquad \qquad  +ie^{-\phi}(2|t_3|^2+24|h_1|^2-|e^{-A}\mu|^2)z\wedge\bar{z}\wedge\hat{\omega}_R \, . \nonumber
\eea

\vspace{0.3cm}

Notice that the parameters in the previous equations have to satisfy further consistency conditions, namely $\d^2 j = \d^2 \hat{\omega}_R = 0$ and $ \d^2 \tilde{j}_i = 0$. More precisely, taking the exterior derivative of \eqref{SU2torsionspar} and \eqref{dertrip} we obtain (the consistency conditions on $\tilde{j}_1$ and $\tilde{j}_2$ give the same equations) 
\begin{align}
 \Re(h_4\bar{a}_2  + 2 h_1 \bar{t}_3) -\Im(e^{-A}\mu t_3) =&0 \label{Finalsystem2} \, , \\
 \Re(e^{-A}\mu h_4)+ \Im(2 \bar{h}_1 h_4+a_2 \bar{t}_3)=&0\label{Finalsystem3} \, . 
\end{align}

\vspace{0.3cm}

In summary, in order to find a generic  $\mathcal{N}=1$ AdS$_4$ vacua with the choice of O-plane of
Table \ref{table:IIBOplanes},  one has to solve \eqref{Finalsystem1}, 
\eqref{Finalsystem2} and \eqref{Finalsystem3}.  The fluxes and the geometry are then given by \eqref{SU2par} and \eqref{fluxpar}.
The general solutions to these equations are easy to obtain but, since the  expressions are not very illuminating, we do not give them in the paper.

\section{Scale separation}
\label{scalesep} 

A question relevant for both compactifications and holography is whether genuine  4-dimensional vacua exist within  10$d$ supergravity.
To this extent some conditions have to be fulfilled:  the string coupling constant $\e^{\phi}$ needs to be tunable small in order to suppress string loop corrections,
for $\alpha^\prime$ corrections to be small the internal volume needs to be tunable large (in string units) and the AdS scale $\Lambda_{AdS}$ needs to be tunable small. 
Moreover to be able to decouple the massive KK modes and to reduce to a fully 4-dimensional theory, 
all of these three conditions must combine in such a way that 
 the AdS length scale, $L_{AdS}$, is parametrically larger than the length scale set by the compact dimensions $L_{KK}$
\beq
\label{scalesep0}
 \frac{ L_{KK} }{L_{AdS}} << 1 \, . 
\eeq

Let us first discuss how to define $L_{KK}$ and $L_{AdS}$.  The four dimensional length scale is set by the inverse of the AdS$_4$ cosmological
constant in the four-dimensional Einstein frame.  We  follow the notation of \cite{Hertzberg:2007wc}. Since all solutions we will consider have constant warp
factor, we will set $e^A=1$. Then we rewrite the  10-dimensional string frame metric  \eqref{10dmetric} as 
\be\label{KKAnsatz}
\d s^2_{10} = \tau_0^2 \tau^{-2}\d s_4^2 + \rho  \, \d \tilde s_6^2\,,
\ee
where $\d s_4^2$ is the 4-dimensional Einstein frame metric. We have rescaled the internal metric 
\beq
\d s_6^2  = \rho  \,  \d \tilde s_6^2 \, ,
\eeq
in such a way that  the modulus $\rho = (\det g_6)^{1/6}$  measures the string frame volume of the internal manifold
and 
\be
\int_6 \sqrt{\tilde g_6} =\mathcal{O}(1)\, .
\ee

The variable $\tau$  is the 4-dimensional dilaton and is given by
\be
\tau^2 =\e^{-2\phi}\rho^3\,.
\ee
With $\tau_0$ we denote the VEV of $\tau$, such that in equation (\ref{KKAnsatz}) only the dynamical part of $\tau$ is used to obtain 4$d$ Einstein frame. 

Then direct dimensional reduction of the 10-dimensional string frame action gives the  4$d$ Planck mass in terms of the  string mass scale $M^2_s$  
\be
m^2_p = \tau_0^2 M^2_s\,.
\ee
We define the dimensionally reduced action as 
\be
S= \int \sqrt g (m^2_p  R - V)\, ,
\ee
such that the scalar potential is a dimension four operator. The AdS cosmological constant  is then defined as
\be
\Lambda_{AdS} = \frac{V}{m_p^2}\,.
\ee
The number $|\mu|^2$ that appears in the supersymmetry equations of the previous section is related to the cosmological constant in the following way
\be
|\mu|^2 =- 6 \Lambda_{AdS}\,.
\ee
We define the AdS length scale as $L^2_{AdS}=-\Lambda^{-1}_{AdS}$,  such that it determines the 4$d$ curvature as follows
\beq
2R^{(4)}_{\mu\nu} = L_{AdS}^{-2}g^{(4)}_{\mu\nu} \, .
\eeq

\vspace{0.3cm}

The size of the internal manifold is less straightforward to define.  A natural guess for the KK scale, which we will adopt in this paper,  is 
\be
L^2_{KK} =\rho\,.
\ee

\vspace{0.3cm}

The  proper way to check the  condition \eqref{scalesep0} would be to compute the  Kaluza-Klein spectrum and see that the masses are indeed much
larger than the AdS scale. Since this is often  not  easy to perform,  one has to rely on some simpler estimates this  ratio. 

\vspace{0.3cm}
A way to determine under which conditions scale separation can be achieved is to study the dependence of the effective four-dimensional potential on two moduli, namely the volume of the compact manifold, $\rho$, and the dilaton, $\phi$.  
As an example we briefly recall how the this can be applied to the  model of   \cite{DeWolfe:2005uu}, which is one of the first constructions of a type IIA vacuum
admitting full moduli stabilisation and scale separation.  The model of  \cite{DeWolfe:2005uu}  corresponds to a compactification
on an orbifold of $T^6$ with  non-zero $F_0, F_4, H$-fluxes and O6 sources. 
The scalar potential depends on the moduli $\tau$ and $\rho$ schematically as 
\be
\label{IIAeffpot}
V(\rho, \tau)/m_p^4 = |H|^2 \rho^{-3}\tau^{-2} + T_{\rm O6}\tau^{-3} + |F_0|^2 \tau^{-4}\rho^3 + |F_4|^2\tau^{-4}\rho^{-1}\,,
\ee
where  $T_{\rm O6}$ is the O6 tension and  is  the only negative term  in the potential.  The coefficients in the potential are a priori functions of the other moduli.  In the particular example of \cite{DeWolfe:2005uu}, 
it can be shown that, while the $H$ and $F_0$ flux are constrained by the tadpole condition to be order one,  the $F_4$ flux is an unbounded flux quantum.  
In what follows we assume that $|H|^2, T_{O6}, |F_0|^2$ are all order one in proper units and $|F_4|^2$ scales as $N^2$, where $N$ is unbounded. From the detailed balance condition (all terms are of the same order in the potential) we can derive the $N$-dependence for $\rho$ and $\tau$ at a critical point
\be
\rho \sim N^{\tfrac{1}{2}}\,,\qquad  \qquad  \tau\sim N^{\tfrac{3}{2}} \qquad ( \e^{\phi}\sim N^{-\tfrac{3}{4}} ) \,, 
\ee
from which we see that large $N$  implies large volume and weak coupling. Secondly, we find that the AdS scale becomes tunably small in the same limit.  Using  the
scaling of the potential and the 4$d$ Planck mass we find
\be 
V/m_p^4\sim N^{-\tfrac{9}{2}} \qquad m_p^2\sim M_s^2\e^{-2\phi}\rho^3 \sim N^3 \qquad L_{AdS}^{-2} \sim N^{-3/2}\,.
\ee
Since $\rho =  L_{KK}^2$  we indeed find scale separation
\be\label{separation}
\frac{L_{KK}}{L_{AdS}}\rightarrow 0\,.
\ee

\vspace{0.3cm}

A similar argument can be also given for IIB solutions.  
On the type IIB side,  an explicit example with the right properties of  tunably large volume, 
small coupling and small AdS scale was found in \cite{Caviezel:2008ik} by T-dualising the type IIA torus example.  A systematic study, from a 4$d$ point of view, of IIB solutions was initiated in \cite{Caviezel:2009tu}.  Typically we have models  with  $F_1, F_3, F_5$ flux, O5, O7 sources on some curved internal manifold. 
The scalar potential can be written as
\be
V(\rho, \tau)/m_p^4 = \tau^{-4} \Bigl( |F_5|^2\rho^{-2} + |F_3|^2 + |F_1|^2\rho^2 \Bigr)  + \tau^{-3}\Bigl(T_{O7}\rho^{\tfrac{1}{2}}  + T_{O5} \rho^{-\tfrac{1}{2}}\Bigr) + R_6 \tau^{-2}\rho^{-1}\,.
\ee
where, as in the type IIA case, the coefficients are functions of all other moduli. The models of \cite{Caviezel:2008ik, Caviezel:2009tu} are characterised by
 two unbounded flux quanta:  $F_5$ and a component of $F_1$,  whereas another component is determined by the O7 planes and not tunable. 
 So let us scale both fluxes
\be
|F_5|^2 \sim N^2\qquad |F_1|^2 \sim N^C\,,
\ee
where $C$ is some positive number. If we then balance $F_5^2$ against $F_1^2$ and $T_{\rm O7}$ we find the following $N$-dependence
\be
\rho\sim N^{\tfrac{1}{2}-\tfrac{C}{4}}\,,\qquad \e^{\phi}\sim N^{-\tfrac{C}{4}}\,. 
\ee
If $0<C<2$  the solution is indeed at large volume and weak coupling for large $N$. The $F_5, F_1, T_{\rm O7}$ contributions, which  set the size of the AdS solution, scale as
\be
V/m_p^4\sim N^{-2 -2C}\,,
\ee
and go to zero at large $N$. If we compute $L^2_{AdS}$ we again find a separation of scales. 
This argument relied on a detailed balance condition for the $F_5, F_1$ and $T_{\rm O7}$ contributions. We have not discussed the other contributions to the scalar potential. In the explicit solutions we derive in this paper all terms in the potential will be of the same order of magnitude.

\vspace{0.3cm}

The Ricci scalar of the internal space 
scales as the inverse metric/ $ \rho^{-1}$. Therefore one would be tempted to conclude that in the large volume limit the two definitions of scale separation 
\begin{align}
& \text{scale separation} \,\,(1) :\qquad\frac{L^2_{AdS}}{L^2_{KK}}\rightarrow \infty\,, \label{scalesep1}\\
& \text{scale separation} \,\,(2) :\qquad \frac{R_6}{R_4} \rightarrow \infty \label{scalesep2}\, ,
\end{align}
are equivalent. 
However, these two definitions do not need to coincide if the normalised curvature $\tilde{R}$
\be
 R_6=\rho^{-1}\tilde{R}(\phi_I)\,,
\ee
is not kept constant  when the limit of large $\rho$ is taken. The other moduli $\phi_I$ that appear inside the normalised  curvature
could also introduce an extra scaling. Below we find an explicit examples for which there is scale separation according to the first definition, but not according to the second definition. Nonetheless that, as we will show in the next sections,  the ratio of the Ricci scalars can be  very useful in setting 
general conditions on the torsion classes of the internal manifold in order to achieve separation of scales. 

\vspace{0.3cm}

\subsection{Separation of scales without sources?}

The most trustworthy solutions are those without any orientifold or D-brane sources since there is no reason to  worry about the smearing approximation or charge quantisation. Even in the case one knows the localised solutions one could have rightful worries about the use of supergravity in the presence of singular sources. A priori, sourceless AdS SUSY vacua can exist  both for $\SU(3)$ structure manifolds
IIA and  $\SU(2)$ structure ones  in IIB. When SUSY is broken many more solutions can exist, see for instance reference \cite{Koerber:2010rn} for the IIA case. It is not clear whether solutions without sources allow for
separation of scales \cite{Tsimpis:2012tu}. The above scaling arguments do not obviously use the presence of a source term. We did include it in the analysis, but we could have equally discarded it. It turns out that it depends on the details of the manifold whether there exists the specific large flux limits that achieve scale separation.  
While no no-go theorem has been found so far,  there is no example known of a solution in 10$d$ SUGRA, whether SUSY or not, that achieves scale separation without sources. 

\vspace{0.3cm}

In the following we give a simple argument that seems to suggest that AdS  without sources do not allow for scale separation. The argument below holds under
two assumptions: 1) there is no warping and 2) the size of the internal manifold cannot be decoupled from its curvature radius. 

Consider a general compactification with RR fluxes $F_p$, $H$ flux and no sources. The scalar potential can be written as
\be
V(\rho, \tau)/m_p^4 =  - R_6 \tau^{-2}\rho^{-1} + |H|^2\rho^{-3}\tau^{-2} +\sum_p |F_p|^2\rho^{3-p}\tau^{-4}\,.
\ee 

The vacua of the theory must be extrema of the scalar potential. One can easily verify that the equations 
\be
\partial_{\rho} V =0\,,\qquad \partial_{\tau} V= 0 
\ee
are specific linear combinations of the dilaton equation of motion in 10 dimensions, the trace over the internal indices of the (trace reversed) Einstein equation, and the external Einstein equation \cite{Danielsson:2009ff}. Upon eliminating $|H|^2$ in terms of the RR field strength densities, these are equivalent to 
\begin{align}
& R_4=2V = -2\sum_p  |F_p|^2 < 0\,,\\
& R_6 = \sum_p \frac{9-p}{2} |F_p|^2>0\, , 
\end{align}
where we did not write down the explicit $\rho, \tau$ dependence anymore. The first condition is an alternative derivation of the Maldacena--Nunez no-go theorem \cite{Maldacena:2000mw} in the simple case of no warping. The second condition was found before \cite{Douglas:2010rt}. With the above equations we can compute the ratio 
\be
r = |\frac{R_6}{R_4}|\,,
\ee
and define scale separation as the possibility to have $r>>1$ (\ref{scalesep2}). However, our equations imply that  $r$ is bounded from above by a number $r_{max}$, since $p<9$. We compute $ r_{max} $ by rewriting the inequality:
\be
\sum_p (\frac{9-p}{2} - 2  r_{max}) |F_p|^2 < 0\,.
\ee
From this one deduces that 
\be
r_{max} = \frac{9-p_{max}}{4}\,,
\ee
where $p_{max}$ is the highest rank field strength that is turned on in the vacuum solution. We then conclude that, under the assumptions we discussed above,  AdS vacua not supported by sources cannot achieve scale separation. 
Note that indeed Freund--Rubin vacua \cite{Freund:1980xh} have $r$ of order one.
It would be most interesting to see whether the same  argument holds when  allowing  for warping.

\section{Explicit examples}
\label{examples}

Natural candidates for our search for  vacua are 
 manifolds for which we can compute the $\SU(2)$ structure explicitly. These are typically coset and group manifolds, although recently progress has been made for non-homogeneous manifolds \cite{Larfors:2010wb, Dabholkar:2013qz}.  In this paper we  study in details the class of coset manifolds  discussed in \cite{Caviezel:2009tu} and nilmanifolds. Our aim is to compute the 10$d$ solutions when they exist and to demonstrate when solutions cannot exist. 
 This was an open issue in \cite{Caviezel:2009tu} and with the pure spinor technology this is quite straightforward to settle.

\vspace{0.3cm}

Group manifolds are particularly simple to analyse because of  the existence of left-invariant forms, which can be used as an expansion basis for the various fields
on the manifold.  In particular, one can  also take the forms defining the $\SU(2)$ structure to be left-invariant. 
Let us recall, for simplicity, that the most general ansatz for the $\SU(2)$ structure forms and the anti-self dual triplet $\tilde{j}_1$, $\tilde{j}_2$ and $\tilde{j}_3$ compatible with the orientifold projections of   Table \ref{table:IIBOplanes} is 
\bea
&& z= z_1e^1+z_2e^2\,, \nonumber\\
&& j=j_1e^{36}+j_2 e^{45}\,,\nonumber\\
&& \omega_R=\frac{j_1j_2}{\omega_{1}}e^{34} + \omega_{1}e^{56}\,,\nonumber \\
&&\omega_I=-\frac{j_1j_2}{\omega_{2}}e^{35}+\omega_{2}e^{46}\,, \label{ansatzSU2a}
\eea
with $z_1 = z_{1R}+i z_{1I}$, $z_2 = z_{2R}+i z_{2I}$ and 
\bea
&& \tilde{j}_1 = j_1e^{36} - j_2 e^{45}  \, , \nonumber\\
&& \tilde{j}_2  = -  \frac{j_1j_2}{\omega_{1}}e^{34} + \omega_{1}e^{56}  \, ,\nonumber  \\
&& \tilde{j}_3  = -\frac{j_1j_2}{\omega_{2}}e^{35} - \omega_{2}e^{46} \, .  \label{ansatzSU2b} 
\eea
The NS and RR fluxes are given by \eqref{fluxpar} 
\bea
\label{fluxparsol}
&&  H=  2 h_1 z \wedge  \hat{\omega}_R + h_4 z \wedge  \tilde{j}_2 +\mathrm{c.c.}\,,\nn\\
&& F_1=  e^{- \phi} (4 i h_1 -\frac{1}{2}  \bar{\mu} e^{-A} )   z +\mathrm{c.c.}\,,\nn\\
&& F_3= - \frac{1}{2} e^{- \phi} (i  \bar{\mu} e^{-A} + 4  h_1)  z \wedge j + \frac{i}{2}  e^{- \phi}  t_3 z \wedge \tilde{j}_1  +\mathrm{c.c.}\,,\nn\\
&& F_5=  e^{- \phi} \, f_5 z \wedge j \wedge j +\mathrm{c.c.}
\eea
For left-invariant forms,  all the exterior derivatives  are given in terms of the structure constants and the  supersymmetry constraints  \eqref{SU2torsionspar} reduce to algebraic equations. 
The computation then proceeds by enforcing  all the algebraic conditions, those coming from  supersymmetry, the Bianchi identities  \eqref{Finalsystem1}  and  the consistency conditions \eqref{Finalsystem2}, \eqref{Finalsystem3}. Finally  \eqref{RRBIs} is used to determine the source terms.

\vspace{0.3cm}

We can use \eqref{normps} to derive the volume of the internal manifold
\beq
\label{6dvol}
{\rm vol}_6 =  \frac{i}{4}z\wedge\bar{z}\wedge j\wedge j |_{\rm top} = j_1j_2\Im(z_1\bar{z}_2) \,,
\eeq
and its metric  (see \cite{Grana:2006kf})

\begin{align}
 g_{ij}=\left(\begin{tabular}{c c c c c c}
 $z_{r1}^2+z_{i1}^2$ & $z_{r1}z_{r2}+z_{i1}z_{i2}$ & 0 & 0 & 0 & 0\\
 $z_{r1}z_{r2}+z_{i1}z_{i2}$ & $z_{r2}^2+z_{i2}^2$ & 0 & 0 & 0 & 0\\
 0 & 0 & $-\frac{ j_1^2 j_2}{\omega_{1}\omega_{2}}$ & 0 & 0 & 0\\
 0 & 0 & 0 & $-\frac{ j_2 \omega_{2}}{\omega_{1}}$ & 0 & 0\\
 0 & 0 & 0 & 0 & $-\frac{ j_2 \omega_{1}}{\omega_{2}}$ & 0\\
 0 & 0 & 0 & 0 & 0 & $-\frac{\omega_{1} \omega_{2}}{j_2}$
\end{tabular}\right)\,.
\end{align}

\vspace{0.3cm}

For this class of models, the Ricci scalar takes  a very simple form when expressed in terms of the $\SU(2)$ torsion classes
\beq
\label{Riccistxt}
R_6 = -4(|S_1|^2 +  4 |S_5|^2 + \Im (S_5 \bar{S}_1) + |T_2|^2 + |T_3|^2 )\,.
\eeq

\vspace{0.3cm}

We are also interested in the total charges associated with the sources.  These can be obtained form \eqref{RRBIs} writing the Bianchi identities  \eqref{BIRRsmeareds} as\footnote{A derivation of this expression can be found in Appendix \ref{riccisapp}.} 
\bea
\label{BIRRex}
&& \d F_1  =  \sum_i N^i_{O7}  \eta_2^i  \nn \\
&&  \d F_3 - H \wedge F_{1} =  \sum_i N^i_{O5}  \eta_4^i \, ,
\eea
where $\eta_2^i$ and $\eta_4^i$ are decomposable two- and four-forms  Poincar\'e dual to the cycles wrapped by the sources.
Deriving \eqref{fluxparsol}, it is straightforward to check that  the O7-plane charges are 
\begin{align}
N_{O7}^{(1)} & = 2  e^{ - \phi}  \frac{j_1 j_2}{\omega_2}  [|\mu|^2  + 8 \, {\rm Im} (\mu h_1) ]\,,\nn\\ 
 N_{O7}^{(2)}& =-2  e^{ - \phi}  \omega_2 [|\mu|^2  + 8 \, {\rm Im} ( \mu h_1) ] \, , 
\end{align}
 while for the O5 we obtain
\begin{align}
N_{O5}^{(1)}=& 2 e^{-\phi}  \, \frac{j_1 j_2}{\omega_1} \,  \Im( z_1 \bar{z}_2) \, \left[2|t_3|^2-|\mu|^2+24 |h_1|^2-2\Im(\mu h_4)-12\Re(h_4 \bar{h}_1) + 2\Re(a_2 \bar{t}_3)\right]\,,\nn\\
N_{O5}^{(2)} =& 2 e^{-\phi} \omega_1 \Im( z_1 \bar{z}_2) \,  \left[2|t_3|^2- |\mu|^2+24 |h_1|^2+2\Im(\mu h_4)+12\Re(h_4 \bar{h}_1)- 2\Re(a_2 \bar{t}_3)\right] \, .
\end{align}

\vspace{0.4cm}

Before moving to the discussion of the explicit examples we summarise in Table  \ref{table3}  whether  
our solutions admit weak coupling, large volume and scale separation.  More precisely we look at possible scalings where,
taking the limit  of small  cosmological constant
\be
|\mu|^2 \rightarrow 0\, ,
\ee
we can have small coupling and large volume.  As can be seen from Table  \ref{table3}  we have also checked whether separation  of scales
according  to the two definitions (\ref{scalesep1}, \ref{scalesep2}) is possible. 
\begin{table}[ht] 
\begin{center}
  \begin{tabular}{ |c || c | c | c | c| }
    \hline
    Manifold                                   & weak coupling & Large volume & scale separation (1) & scale separation (2)\\ \hline\hline
     $\frac{\SU(3)}{\SU(2)}\times \U(1)$        & $ \times$           & \checkmark           &  \checkmark          &  $\times$      \\ \hline
     $\frac{\SU(2)^2}{\U(1)}\times \U(1)$        &  $\times $          & \checkmark   &  \checkmark          &  $\times$        \\ \hline   
     Nil  3.14        &  \checkmark           &  $\times$      &  \checkmark          &   \checkmark  \\ \hline   
     Nil  4.1        &  \checkmark           & \checkmark   &  \checkmark          &  \checkmark        \\ \hline   
     Nil  5.1        &  \checkmark           & \checkmark   &  \checkmark          & $\times$         \\ \hline   
  \end{tabular}
\caption{The scaling regimes for the various manifolds with SUSY AdS vacua.}
\label{table3}
\end{center}
\end{table}
Only the solutions on Nil 4.1 and 5.1 can be tuned into a trustworthy regime. The solution on Nil 3.14 cannot be achieved for large volume and furthermore suffers from having a singular limit (vanishing volume) if scale separation is required. 
We also notice that for several examples there is no match between the two criteria for scale separation. 

In finding the appropriate limits we have taken a conservative and safe viewpoint where each source term was taken to have finite prefactors as a consequence of charge quantisation (see footnote below equation (\ref{BIRRsmeareds})). Since not all source terms are represented by forms in the cohomology of the internal space, it is possible that charge quantisation is less restrictive on such forms and that certain numbers do not need to take fixed values. In that case it is possible that certain solutions do allow weak coupling and scale separation although the above table indicates otherwise. We leave such subtle issues for further investigation.

\subsection{Coset manifolds}

We refer to  \cite{Koerber:2008rx} for a thorough discussion of coset manifolds and $G$-structures.  Here we simply recall some simple facts that help making our 
derivation clearer.

A coset manifold $M = G/H$ where $G$ is a Lie group and $H$ is a closed subgroup of $G$, is  completely determined by
the corresponding  algebrae, $g$ and $h$.  We denote by $\{H_a\}$, with $a = 1, \dots \, \dim H$, a basis of generators of $h$ and by $ \{ K_i \}$, with $i = 1, \ldots, \dim G - \dim H$
a basis for the complement of $h$ in $g$. Then the structure constants are given by
\bea
\label{genalgebra}
&& [H_a ,  H_b] = f^c_{a b} H_c \, , \nonumber\\
&& [H_a ,  K_i] = f^j_{a i} K_j  + f^c_{a i} H_c \, , \nonumber\\
&& [K_i, K_j] = f^i_{jk} K_i + f^a_{jk} H_a \, .
\eea
The coframe $e^i(y)$ on $G/H$ is defined by
\beq
L^{-1} \d L = e^i K_i + \omega^a H_a \, , 
\eeq
where $L(y)$ is a coset representative and $y^i$ are local coordinates on $G/H$.  
A $p$-form 
\beq
\phi = \phi_{i_i \ldots i_p} e^{i_1} \wedge \ldots \wedge e^{i_p}
\eeq
is then said to be left-invariant under the action of $G$  if and only if its coefficients $ \phi_{i_i \ldots i_p}$ are constant and
\beq
\label{leftinvc}
f^j_{a[i_1}\phi_{i_2 \dots i_p] j} = 0 \, .
\eeq
From the algebra \eqref{genalgebra} we have 
\beq
\d e^i =  -\frac{1}{2} f^i_{j k} e^j \wedge e^k  - f^i_{a j} \omega^a \wedge e^j \, .
\eeq
It is  then easy to show that \eqref{leftinvc} guarantees that the exterior derivative preserve the property of left-invariance.  

\vspace{0.3cm}

As mentioned before, we want the $\SU(2)$ structure to be also left-invariant. As shown in  \cite{Koerber:2008rx}  this requires  that $H \in \SU(2)$.
The list of reductive coset manifold that satisfy this property is \cite{Caviezel:2009tu}
\beq
\frac{\SU(3)\times U(1)}{\SU(2)} \qquad  \frac{\SU(2)^2}{U(1)}\times U(1) \qquad \SU(2)\times \SU(2) \qquad \SU(2)\times U(1)^3 \, .
\eeq
The rest of this section is devoted to the study of $\mathcal{N}=1$ SUSY  AdS$_4$ on such manifolds.


\subsubsection{$\frac{\SU(3)\times U(1)}{\SU(2)}$}

Of the 9 generators of $\SU(3) \times U(1)$ we denote by $T_2$ and $T_7, T_8, T_9$ the  generators of  $U(1)$ and $\SU(2)$,  respectively.
The algebra is given by
\begin{align}
& f_{46}^1 =-\frac{\sqrt3}{2} \quad (\text{and cyclic})\,,\qquad f^1_{35} =\frac{\sqrt3}{2} \quad (\text{and cyclic})\,,\nonumber\\
& f_{78}^9 = 1 \quad (\text{and cyclic})\,,\nonumber \\
& f^7_{65}= f^7_{34} = f^8_{63}= f^8_{45}= f^9_{64}= f^9_{53} =\frac{1}{2}\quad (\text{and cyclic})\,.\label{algebra}
\end{align}
The left-invariant forms compatible with the O5 and O7 projections are 
\begin{center}
\begin{tabular}{lcl} 
1 forms & \quad & $e^1$, $e^2$ \, ,\\
2 forms & \quad & $e^{36} + e^{45}$, $e^{34} + e^{56}$, $e^{35} - e^{46} \, , $
\end{tabular}
\end{center}

Since the $\SU(2)$ structure forms must also be left-invariant,  in the ansatz \eqref{ansatzSU2a} and \eqref{ansatzSU2b} we set 
\beq
j_1 = j_2 \qquad  \omega_1 = \epsilon_1 j_2 \qquad \omega_2=\epsilon_2 j_2 \, , 
\eeq
with $\epsilon_1 = \pm 1 $ and $\epsilon_2=\pm 1$.  It is easy to see that none of the forms $\tilde{j}_i$ is left-invariant,  which implies $t_3= h_4 = 0$.  Solving the constraints \eqref{Finalsystem1},
\eqref{Finalsystem2} and \eqref{Finalsystem3} gives the solution 
\begin{align}
 h_1=& \frac{i}{2} \frac{\Re(\mu\bar{z}_2)}{z_2} \,,\nonumber\\
z_1=&- \frac{\epsilon_1}{2} \frac{\sqrt{3}\mu |z_2|^2}{ \Im(\mu\bar{z}_2)^2}\,,\nonumber\\
j_2=&-  \epsilon_1 \epsilon_2 \frac{3}{8} \frac{|z_2|^2}{\Im(\mu\bar{z}_2)^2}\,,\nonumber\\
a_2=&0\,,
\end{align}
where
\begin{align}
\rho^3=&- \epsilon_1 \frac{9\sqrt{3}}{128}  \frac{|z_2|^6}{\Im(\mu\bar{z}_2)^5} \,,\nonumber\\
 R_6=&- 4 \Big[ |\mu |^2 -2 \frac{\Im(\mu  \bar{z}_2)^2}{|z_2|^2} \Big]\,,
\end{align}
and the orientifold charges are
\begin{align} 
 N_{O7}^{(1)}=& N_{O7}^{(2)}=-  \frac{3 \epsilon_1}{4} \,e^{- \phi} \,  \left[ 1 + 5 \,  \frac{ \Re(\mu \bar{z_2})^2}{\Im(\mu\bar{z}_2)^2} \right] \,,\nonumber\\
N_{O5}^{(1)}=& N_{O5}^{(2)}= \frac{3\sqrt{3}}{8} \epsilon_1\epsilon_2 e^{-\phi} \, |z_2|^2
\left[-\frac{1}{\Im(\mu\bar{z}_2)} + 5 \, \frac{ \Re(\mu \bar{z}_2)^2}{\Im(\mu\bar{z}_2)^3} \right] \,.
\end{align}

Note that by consistency the orientifold planes should wrap directions whose dual source forms should be left-invariant. The source forms that are Poincare dual to the surfaces wrapped by the O5 planes are not left-invariant, although we have said O planes have to be consistent with the left-invariant forms. This problem is however cured since the sum of the two O5 forms, $e^{1234}+e^{1256}$  is left-invariant. If we interpret each of these two terms as a separate orientifold source with its own involution then consistency requires the two O-planes to have exactly the same charge, such that the source term in the Bianchi identity for $F_3$ is given by the sum of the two forms, and hence left-invariant. This is anyhow a necessary requirement from the point of view of charge quantisation. Orientifolds, unlike D-branes, cannot be stacked. So for each involution we should have a single unit of orientifold charge.

\vspace{0.3cm}

\subsubsection{$\frac{\SU(2)^2}{\U(1)}\times \U(1)$}

For this coset, out of the 7 generators of $\SU(2)^2 \times \U(1)$ we denote by $T_7$ the  generator of the $U(1)$. The algebra is given by :
\begin{align}
 &f^1_{35}=1~\mathrm{(and~cyclic)},~~~f^7_{46}=1~\mathrm{(and~cyclic)}\nonumber\\
&f^1_{46}=f^5_{37}=-f^3_{57}=-1 \, .
\end{align}

As in the previous case, the $\SU(2)$ structure must be left-invariant. This means that in the ansatz \eqref{ansatzSU2a} and \eqref{ansatzSU2b} we set 
\beq
j_1 = j_2 \qquad  \omega_1 = \epsilon_1 j_2 \, , 
\eeq
with $\epsilon_1=\pm1$. As before, the requirement of left-invariant implies $h_4=0$ and $t_3=0$.
The solution is 
 \begin{align}
z_{1}=&- \epsilon_1 \, \frac{ \mu |z_2|^2}{2 \Im( \mu \bar{z_2} )^2}  \,,\nonumber\\
\omega_2=& -\epsilon_1 \frac{ |z_2|^2} {4 \Im(\mu \bar{z}_2)^2} \,,\nonumber\\
j_2=& -\epsilon_2 \frac{ |z_2|^2} {4 \Im(\mu \bar{z}_2)^2} \,,\nonumber\\
h_1=& \frac{i}{2}  \frac{ \Re(\mu \bar{z_2} )}{z_2} \,,\nonumber\\
h_4=&t_3=0 \,,\nonumber\\
a_2=&i\frac{\Im(\mu\bar{z}_2)}{z_2}\,,
\end{align}
with $\epsilon_2=\pm 1$.
The volume and curvature are
\begin{align}
\rho^3 =&  \frac{\epsilon_1}{32}  \frac{|z_2|^6}{ \Im(\mu \bar{z}_2)^5} \,,\nonumber\\
 R_6=&-4\left[ |\mu |^2  -2 \frac{\Im(\mu \bar{z}_2)^2}{|z_2|^2}\right] \,,
\end{align}
and the orientifold charges are
\begin{align} 
 N_{O7}^{(1)}=& N_{O7}^{(2)}=  -\frac{\epsilon_1}{2}  e^{-\phi} \left[1+5 \, \frac{\Re(\mu \bar{z}_2)}{\Im(\mu \bar{z}_2)} \right] \,,\nonumber\\
N_{O5}^{(1)}=& N_{O5}^{(2)}=  \frac{\epsilon_2}{4} e^{-\phi} |z_2|^2 \left[ -\frac{1}{\Im(\mu \bar{z}_2)} + 5 \,  \frac{ \Re( \mu \bar{z}_2)^2}{ \Im(\mu \bar{z}_2)^3} \right] \,.
\end{align}

\subsubsection{$\SU(2)\times \SU(2)$ and $\SU(2)\times \U(1)^3$}

There are no SUSY solutions on these two manifolds. This is most easily  seen for  $\SU(2)\times \SU(2)$ since the SUSY equations 
(see \eqref{SU2torsionspar})
require the
one-form $\Im(\bar{\mu}z)$  to be closed whereas there are no closed (left-invariant) one-forms on $\SU(2)\times \SU(2)$.

\subsection{Nilmanifolds}

Nilmanifolds have  Lie groups defined by a nilpotent algebra as their covering space.  Having negative curvature they are natural candidates for flux 
compactifications \cite{Grana:2006kf}.  In six dimensions there are
 34 isomorphism classes of simply-connected six-dimensional nilpotent Lie groups. The full list of algebrae can be found in Table 4  of  \cite{Grana:2006kf} and in \cite{Cavalcanti},
 together with the pure spinors  and orientifold projections compatible with each algebra. In particular it is easy to see that very few algebrae 
 are compatible with 
 the orientifold projections we are imposing.  Using the notation of \cite{Grana:2006kf}\footnote{With respect to Table 4 of  \cite{Grana:2006kf} we have relabeled 
 the one-forms in order to match our orientifold projections} they are

\begin{table}[h]

\begin{center}
{\small
\begin{tabular}{|  l|  |l | |c|c|}
\hline
$n$ & Nilmanifold class  &   $b_1$ &  $b_2$ \\
\hline
\hline
3.3 & $(0, 35, 0, 0, 13, 14)$ & 3  & 6   \\

\hline
3.13 & $(-35  - 46, 0,0,0,23, -24)$  &   3 & 5 \\

\hline
3.14 & $(-35 + 46, 0,0,0,23,-24)$ &  3 & 5  \\

\hline
\hline

4.1 & $(-35 - 46, 0,-25,0,0,0)$ & 4 & 6 \\

\hline
4.2 & $(-35,0,-25,0,0,0)$  & 4  & 7 \\

\hline

4.5 & $(-46,0,25,0,0, 0) $ & 4  &  8 \\
\hline

4.6 & $(0,0,0,0, 13, 14)$  &  4  & 9 \\

\hline \hline
5.1 &  $(35+46,0,0,0,0,0)$  &   5   & 9   \\
\hline

5.2 & $(35,0,0,0,0,0)$ &  5  & 11 \\
\hline
\hline

6.1 & $(0,0,0,0,0,0)$ &  6  & 15 \\
\hline

\end{tabular} }
\end{center}
\caption{\label{ta:nil} Six--dimensional nilmanifolds compatible with the O5/O7 projections.
}
\end{table}

Most of the algebrae above are ruled out by the first SUSY condition in \eqref{SU2torsionspar}
\beq
\d z = 2 \mu  \hat{\omega}_I  \, .
\eeq
Indeed using the definition of the $\SU(2)$ structure \eqref{ansatzSU2a}, it easy to see that only the algebrae  $n 3.13$, $n 3.14$,  $n 4.1$ and $n 5.1$ have structure constants suitable to satisfy this condition.
In the rest of this section we give the $\mathcal{N}=1$ AdS$_4$ solutions for the manifolds $n 3.14$,  $n 4.1$ and $n 5.1$ ($n3.13$ is almost 
identical to $n3.14$).  Notice, that differently from the solutions on coset manifolds, these
vacua can have non trivial torsion $t_3$ and H-flux $h_4$. 

On a nilmanifold all forms are left-invariant.  To our purposes this means that the solutions are less constrained then for coset manifolds.
In particular we have no constraints on the moduli in the $\SU(2)$ structure. For this reason we expect this vacua to have generically some
unfixed moduli.

\subsubsection{Model 3.14}
The solution is given by
\begin{align}
z_1=&-2\mu \epsilon_1 \sqrt{j_1j_2}\,, \nonumber\\
\omega_2=&\epsilon_1\sqrt{j_1j_2}\,, \nonumber\\
h_1=&\frac{i}{4}\bar{\mu}\left(2+\frac{i(j_1+j_2)\omega_1}{2\Im(\mu\bar{z}_2)j_1j_2}\right)\,, \nonumber\\
h_4=&\frac{\bar{\mu}(j_1+j_2)\omega_1}{4\Im(\mu\bar{z}_2)j_1j_2}\,, \nonumber\\
t_3=&-a_2=\frac{i\bar{\mu}(j_1-j_2)\omega_1}{4\Im(\mu\bar{z}_2)j_1j_2}\,.
\end{align}
where the volume and curvature read
\begin{align}
\rho^3=& 2\epsilon_1\Im(\mu\bar{z}_2)(j_1j_2)^{\frac{3}{2}}\,, \nonumber\\
 R_6=&-2|\mu|^2\left(2+\frac{(j_1^2+j_2^2)\omega_1^2}{4\Im(\mu\bar{z}_2)^2j_1^2j_2^2}\right)\,.
\end{align}
The orientifold charges are :
\begin{align} 
 N_{O7}^{(1)}=& N_{O7}^{(2)}=-10\epsilon_1e^{-\phi}|\mu|^2\sqrt{j_1j_2}\,, \nonumber\\
N_{O5}^{(1)}=&-\epsilon_1\frac{4e^{-\phi}|\mu|^2}{\sqrt{j_1j_2}\Im(\mu\bar{z}_2)\omega_1}\left(5(j_1j_2\Im(\mu\bar{z}_2))^2+(j_1^2+j_1j_2+j_2^2)\omega_1^2\right)\,,\nonumber\\
 N_{O5}^{(2)}=&-20 e^{-\phi}\epsilon_1|\mu|^2\sqrt{j_1j_2}\Im(\mu\bar{z}_2)\omega_1\,.
\end{align}

\vspace{0.3cm}

\subsubsection{Model 4.1}

The solution is 
\begin{align}
z_1=&-2\mu\epsilon_1\sqrt{-j_1j_2}\,,\nonumber\\
\omega_2=&-\epsilon_1\sqrt{-j_1j_2}\,,\nonumber\\
h_1=&\frac{i}{4}\bar{\mu}\left(2+\frac{ij_1}{2\Im(\mu\bar{z}_2)\omega_1}\right)\,,\nonumber\\
h_4=&i a_2=i t_3=-\frac{\bar{\mu}j_1}{4\Im(\mu\bar{z}_2)\omega_1}\,.
\end{align}
where
\begin{align}
\rho^3=&-2\epsilon_1\Im(\mu\bar{z}_2)(-j_1j_2)^{\frac{3}{2}}\,,\nonumber\\
 R_6=&-2|\mu|^2\left(2+\frac{j_1^2}{4\Im(\mu\bar{z}_2)^2\omega_1^2}\right)\,,
\end{align}
and the charges are
\begin{align} 
 N_{O7}^{(1)}=&- N_{O7}^{(2)}=-10\epsilon_1e^{-\phi}|\mu|^2\sqrt{-j_1j_2}\,,\nonumber\\
N_{O5}^{(1)}=&20\epsilon_1e^{-\phi}|\mu|^2(-j_1j_2)^{\frac{3}{2}}\frac{\Im(\mu\bar{z}_2)}{\omega_1}\,,\nonumber\\
 N_{O5}^{(2)}=&-4\epsilon_1e^{-\phi}|\mu|^2\sqrt{-j_1j_2}\left(\frac{j_1^2+5\Im(\mu\bar{z}_2)^2\omega_1^2}{\Im(\mu\bar{z}_2)\omega_1}\right)\,.
\end{align}

\vspace{0.3cm}

\subsubsection{Model 5.1}

The solution is :
\begin{align}
 z_1=&-2\mu\epsilon_1\sqrt{-j_1j_2}\,,\nonumber\\
\omega_2=&-\epsilon_1\sqrt{-j_1j_2}\,,\nonumber\\
h_1=&\frac{i\bar{\mu}}{2}\,,\nonumber\\
h_4=&a_2=t_3=0\,,
\end{align}
where
\begin{align}
\rho^3=&-2\epsilon_1\Im(\mu\bar{z}_2)(-j_1j_2)^{\frac{3}{2}}\,,\nonumber\\
R_6=&-4|\mu|^2 \,,
\end{align}
and the charges are
\begin{align} 
N_{O7}^{(1)}=&- N_{O7}^{(2)}= -10\epsilon_1e^{-\phi}|\mu|^2\sqrt{-j_1j_2}\,,\nonumber\\
N_{O5}^{(1)}=&20\epsilon_1e^{-\phi}|\mu|^2(-j_1j_2)^{\frac{3}{2}}\frac{\Im(\mu\bar{z}_2)}{\omega_1} \,,\nonumber\\
N_{O5}^{(2)}=&-20\epsilon_1e^{-\phi}|\mu|^2\sqrt{-j_1j_2}\Im(\mu\bar{z}_2)\omega_1\,.
\end{align}
This solution can be obtained from a T-duality of the O6 toroidal orientifold in massive IIA SUGRA.


\section{Discussion}

In this paper we  have studied what are the general conditions to have  10-dimensional   $\mathcal{N}=1$ AdS$_4$ vacua from IIB compactifications on smeared O5/O7 orientifolds.  The orientifold projections forces the internal manifold to be of $\SU(2)$ structure.  
We give the supersymmetry conditions both in terms of the pure spinors defined on the internal manifold \cite{Grana:2006kf} 
and in terms of $\SU(2)$ torsion classes. Our analysis is completely 10-dimensional and parallels the existing results for solutions in type IIA \cite{Lust:2004ig}. 

The main purpose of the paper was to look for possible candidates for scale separation in purely classical type IIB backgrounds. 
It is natural to look first for relatively simple classes of manifolds,  whose geometry is under control, namely  reductive cosets and nilmanifolds.
We applied our general formalism to such manifolds and we checked whether the vacua we found admit limits where 
the cosmological constant and the string  coupling are small, while the internal volume is large. We also checked scale separation, using both criteria we 
discussed in Section \ref{scalesep}.  Our results are summarised in Table \ref{table3}. Clearly, in order to have a complete prove that 
scale separation is indeed possible, one should compute the Kaluza--Klein spectrum. It is nonetheless promising that some of the vacua
survive a first analysis based on scaling arguments. It would also be interesting to have a better understanding of when and why
the different scaling criteria have to agree or not.


\vspace{0.3cm}

There are various issues that call for further research. Most pressing is the status of the orientifold sources. Apart from the charge quantisation, which has not been worked out, it is essential to understand how to localise the orientifold planes. Likely this question is more tractable using the pure spinor formalism, as was attempted for localised O6 solutions in massive IIA \cite{Saracco:2012wc}. If an analogy can be made with solutions that feature parallel orientifolds \cite{Blaback:2010sj, Grana:2006kf} then one can expect that localisation will change the geometry but that the very existence of the solution is not invalidated. It should also be possible to find solutions of our type (IIB $\SU(2)$ structures) without any sources, which should be relevant for holography. We have not been able to do this for the class of manifolds we considered (reductive cosets and nil manifolds\footnote{This is obvious for nil manifolds as they are negatively curved whereas source less solutions require positively curved internal spaces \cite{Douglas:2010rt}}.), but it would be interesting to look for instance at Solv manifolds. It is perhaps even more relevant to break supersymmetry and to look for non-SUSY AdS vacua in this context or even dS vacua. In IIA meta-stable non-SUSY AdS vacua have been found using Ansatze that are close to that of the SUSY AdS solutions \cite{Koerber:2010rn}. Interestingly, the same has been done for dS solutions \cite{Danielsson:2009ff, Danielsson:2010bc, Andriot:2010ju, Danielsson:2011au}\footnote{These de Sitter constructions are in part inspired on the proposals in \cite{Silverstein:2007ac, Haque:2008jz}.}, although none of the latter examples turned out meta-stable. Clearly more examples are required and the results in this paper could offer a first step towards achieving this. Already a (unstable) de Sitter critical point was numerically found in \cite{Caviezel:2009tu}. It would be worthwhile to verify whether this numerical solution lifts to a simple 10-dimensional solution.  Finally we hope that some of the $\SU(2)$ structure technology we developed in this paper has its applications to $\SU(2)$ structure solutions in other contexts, such as in IIA, see for instance \cite{Barranco:2013fza}.

\subsection*{Acknowledgements}
We would like to thank   U. Danielsson, A. Tomasiello, D. Tsimpis, T. Wrase and A. Zaffaroni for  useful discussions.
TVR is supported by a Pegasus Marie Curie fellowship of the FWO.  MP would like to thank the Theory Group at Imperial College for the kind hospitality
during the final part of this project. 
\newpage
\appendix

\section{Conditions for AdS$_4$ vacua with $\mathcal{N}=1$ SUSY in type IIB SUGRA}
\label{SUSYcondps}
In this Appendix we derive  the general conditions that  the fluxes and the internal geometry must satisfy in order to have four-dimensional AdS vacua
with  $\mathcal{N}=1$ supersymmetry in type IIB supergravity. 

We consider ten-dimensional geometries that are (warped) products of four-dimensional Anti-de Sitter  space  times a compact six-dimensional internal manifold, $Y$,
and we allow for non-trivial fluxes that do not break the maximal symmetry of AdS$_4$.  We also restrict our attention to vacua with $\mathcal{N}=1$ supersymmetry.
Backgrounds of this kind can be determined  simply by imposing the supersymmetry variations and the Bianchi identities for the fluxes\footnote{For a space-time which is a warped product $M_d \times Y_{10-d}$ it can be shown that the supersymmetry equations plus flux Bianchi identities imply the full set of equations of motions.}.  To analyse the supersymmetry variations we will use the formalism of  Generalised Complex Geometry \cite{Hitchin:2004ut, Gualtieri:2003dx}.  The idea is that the ten-dimensional supersymmetry conditions can be rewritten as a set of differential equations 
on  globally defined forms  on the internal  manifold  \cite{Grana:2004bg}.  

\subsection{$\SU(3)$ and $\SU(2)$ structures}

In this section we present some basic definitions and identities for $\SU(3)$ and  $\SU(2)$ structures with the purpose of fixing conventions and providing the necessary tools to follow the derivations in the paper.

In type IIB  $\mathcal{N}=1$ compactifications to a maximally symmetric four-dimensional manifold the ten-dimensional supersymmetry parameters factorise as
\beq
\label{10dspinors}
\epsilon_i = \zeta_+ \otimes \eta^i_+ + \zeta_- \otimes \eta^i_-  \qquad \qquad i=1,2 \, ,
\eeq
where $\zeta_+$ and $\eta^i_+$ are Weyl spinors in four and six dimensions respectively, and $\eta^i_- = (\eta^i_+)^\ast$ and $\zeta_- = (\zeta_+)^\ast$.

We assume that the spinors $\eta_1$ and $\eta_2$ are globally defined. This means that the structure group of $Y$ is reduced to a subgroup $G \in \SO(6)$.  What
$G$ is depends on the relation between the two spinors, $\eta_i$, which generically  $\eta^1_+$ and $\eta^2_+$ are neither parallel nor orthogonal.  We can 
parametrise them as 
\bea
&& \eta^1_+ = a \eta_+ \,,\nonumber\\
&& \eta^2_+ = b (\kpar \eta_+ + \frac{1}{2} \kper  z_m \gamma^m \eta_-) \,,\label{6dspinors}
\eea
where $z =  z_m \gamma^m$ is a six-dimensional one-form, $\eta_+$ is a globally defined Weyl spinor of norm one  and $\kpar^2 + \kper^2 = 1$. When $\kpar = 1$ and $\kper=0$ the internal manifold is said to be of $\SU(3)$ structure,
in the opposite case $\kpar = 0$ and $\kper=1$ the structure is $\SU(2)$, while the general case is often referred to as dynamical $\SU(2)$  structure. $a$ and $b$ are two complex functions
determining the norm of $\eta^1$ and $\eta^2$
\beq
\eta^{1 \, \dagger} \eta^1 = |a|^2  \qquad \qquad \eta^{2 \, \dagger} \eta^2 = |a|^2 \, .
\eeq

\vspace{0.3cm}

An  equivalent definition of a  $G$-structure is given in terms of invariant forms on the manifold, which can be constructed as bilinears in the internal spinors.
For $\SU(3)$ these are a real two-form and a holomorphic three-form
\bea
\label{SU3strf}
&& J_{mn} =   -i \eta_+^\dagger \gamma_{mn} \eta_+  \nn\,, \\
&& \Omega_{mnp}  =  -i \eta_-^\dagger \gamma_{mnp} \eta_+ \,,
\eea
satisfying
\beq
J \wedge \Omega  = 0\,, \qquad \qquad  J \wedge J \wedge J   = \frac{3}{4} i \, \Omega \wedge \bar{\Omega} \, ,
\eeq
and 
 \begin{align}
 & \ast \Omega = -i \Omega\,,\label{SU31}  \\
 & \ast  J =-\tfrac{1}{2}J^2\, . \label{SU32}  
 \end{align}
The volume of the internal manifold is given by $J^3 = -6 \,   {\rm vol}_6 $.
 
\vspace{0.3cm} 
 
For the $\SU(2)$ structure, we have a complex one-form, a real and a holomorphic two-form
\bea
\label{SU2strf}
&& z_m =  - \chi^\dagger_- \gamma_m \eta_+ \,,\\
&& j_{mn} =  -i \eta^\dagger_ + \gamma_{mn} \eta_+ + i  \chi^\dagger_ + \gamma_{mn} \chi_+ \,,\\
&& \omega_{mn} = - i \chi^\dagger_+ \gamma_{mn} \eta_+\,,
\eea
where  $\chi_+ = \frac{1}{2} z \eta_-$ 
satisfying
\bea
&& z \llcorner \bar{z} = 2\,, \qquad   z \llcorner z =  \bar{z} \llcorner \bar{z} = 0\,, \\
&& j \w \omega =0 \,, \\
&&  z \llcorner j  = z \llcorner \omega = 0 \,, \\
&& j \w j = \frac{1}{2} \omega \w \bar{\omega} \,.
\eea
The one-form $z$ provides an almost product structure on $Y$,  defined  locally by
\beq
\label{products}
R_m^n = z_m \bar{z}^n + \bar{z}_m z^n - \delta_m^n  \,  , \qquad \qquad m,n=1,  \ldots,  6 \, , 
\eeq 
which induces a (global)  decomposition of the tangent space in
\beq 
T M = T_2 M \oplus T_4 M \,. \label{decomposition}
\eeq
Notice that the sub-bundle $T_2 M$ is spanned by the real and imaginary parts of the form $z$. \\

\subsubsection{Torsion classes}
\label{torsions}

The forms defining a $G$-structure are generically not closed. Their differential can be decomposed into representations of the structure group $G$, the so called
torsion classes, and we classify the different $G$-structures depending on which torsion class is non-zero \cite{Gauntlett:2002sc, Gauntlett:2003cy}. 

\vspace{0.3cm}

For $\SU(3)$ structure such decomposition is
very simple
\bea
\label{SU3torsions}
&& \d J  =  \frac{3}{2}  {\rm Im} (\bar{W}_1 \Omega) + W_4 \wedge J + W_3 \, , \nn  \\
&& \d \Omega =  W_1 \wedge J \wedge J + W_2 \wedge J + \bar{W}_5 \wedge \Omega \, ,
\eea
where $W_1$ is a complex scalar which is a singlet of $SU(3)$,  $W_2$ is  a complex primitive (1,1)-form transforming in the adjoint,  
$W_3$ a real primitive (2,1) and (1,2) form in the $\bf{6} \oplus \bar{\bf{6}}$, $W_4$ a real vector and
$W_5$ a  complex (1,0)-form in the $\bf{3} \oplus \bar{\bf{3}}$.

The same expansion for $\SU(2)$ structures is more involved since the torsion classes  are now 20:  8 complex singlet $S_i$, 8 complex doublets $V_i$ and
4 complex triplets $T_i$. We have\footnote{A simple way to deduce such classes is to
decompose the $\SU(3)$ torsion classes according to $\SU(3) \rightarrow \SU(2) \times \U(1)$. We added two vector representations  in $\d z$ that were missing in \cite{Dall'Agata:2003ir}.}
\bea
\label{SU2torsions}
&& \d z = S_1 \omega +S_2j+S_3 z \wedge \bar{z}+S_4\bar{\omega} + z \wedge (V_1+\bar{V}_2)+\bar{z}  \wedge (V_3+\bar{V}_4)+T_1 \,, \nn \\
&& \d j= S_5 \bar{z} \wedge \omega +S_6  z \wedge \omega + \frac{1}{2}(S_7 + \bar{S}_8)  z \wedge j +j \wedge V_5 + z \wedge \bar{z} \wedge V_6
+ z  \wedge T_2+\mathrm{c.c.} \,, \nn \\
&& \d \omega = S_7 z \wedge \omega + S_8 \bar{z} \wedge \omega -2 \bar{S}_{5} z \wedge j -2 
\bar{S}_{6} \bar{z} \wedge j  + i z \wedge \bar{z} \wedge (\bar{V}_6  \llcorner \omega) +j \wedge (V_7+\bar{V}_{8})  \nn \\ 
&& \qquad +  z \wedge T_3 +\bar{z} \wedge T_4  \, ,
\eea
where the relations between the representations in $\d j$ and $\d \omega$ are implied by the conditions $\d(j \wedge \omega) = \d(j \wedge j)=  \d (\omega \wedge \omega) = 0$.

Notice that the doublet $V_i$ are holomorphic vectors with  respect to the complex structure
defined by $j$,  
\beq
 \omega \wedge  V_i= 0 \, 
\eeq
while the $T_i$ are (1,1) and primitive
\beq
j \wedge T_i= \omega \wedge T_i=0 \,,
\eeq
and can be decomposed on the basis of anti-self dual two-forms $\tilde{j}_1$,  $\tilde{j}_2$,  $\tilde{j}_3$ transforming in the ${\bf 3}$  of  $\SU(2)$
\beq
\label{tripletdec}
T_i = \sum_{a=1}^3 t^{i}_a \tilde{j}_a\,.
\eeq

\subsection{Conditions for AdS$_4$ vacua}

$G$-structures have been successfully applied to flux compactifications as tools to classify the internal manifolds allowing for SUSY flux vacua. 
The idea is to rewrite the supersymmetry variations as differential equations on the $G$-invariant forms, then susy solutions are obtained equating fluxes
and torsions in the appropriate representations.  In particular, for four-dimensional flux compactifications, the relevant structures are $\SU(3)$ and $\SU(2)$.
Generalised Complex Geometry \cite{Hitchin:2004ut, Gualtieri:2003dx} is a further refinement of this
approach which allows for a unified treatment of all  flux vacua with a given amount of supersymmetry, encompassing in such a way all possible $G$-structures.

\vspace{0.3cm}

In this paper we focus on $\mathcal{N}=1$ supersymmetry. 
As shown in \cite{Grana:2004bg}, the  ten-dimensional supersymmetry conditions can be rewritten as a set of differential equations on a pair of globally defined
poly-forms  on $Y$.  Such polyforms, or pure spinors,   are constructed by tensoring the two supersymmetry parameters 
on the internal manifold $\eta^1$ and $\eta^2$
\beq
\label{purespapp}
\Phi_\pm  = \eta^1_+ \otimes \eta^{2 \dagger}_\pm\,.
\eeq
The explicit expression for $\Phi_\pm$ depends on the relation between  $\eta^1_+$ and $\eta^2_+$.  In the general case \eqref{6dspinors}, they  are defined as 
\bea
\label{Phimapp}
&& \Phi_-= -\frac{a b}{8} z \wedge (\kper e^{ -ij }  + i \kpar \omega ) \, , \\
\label{Phipapp}
&& \Phi_+=  \frac{ a \bar{b}}{8} \, e^{ z \bar{z} /2} (\kpar e^{-ij}  -i\kper \omega ) \, ,
\eea
where $z$, $j$ and $\omega$ are the $\SU(2)$ structure forms \eqref{SU2strf}.   For the special cases of $\SU(3)$ structure ($\kpar=1, \kper =0 $) and rigid $\SU(2)$ structure 
($\kpar=0, \kper =1$) the pure spinors reduce to 
\bea
\label{SU3ps}
&& \Phi_-= - i \frac{a b}{8}  \Omega  \, , \\
\label{Phip}
&& \Phi_+=  \frac{ a \bar{b}}{8} \,   e^{-iJ}   \, ,
\eea
where $\Omega$ and $J$ are the $\SU(3)$ invariant forms defining the $\SU(3)$ structure, \eqref{SU3strf}, and 
\bea
\label{SU2ps}
&& \Phi_-= -\frac{a b}{8} z \wedge e^{ -ij }  \, , \\
\label{Phip}
&& \Phi_+= -i  \frac{ a \bar{b}}{8} \, e^{ z \bar{z} /2}  \omega  \, , 
\eea
respectively.

\vspace{0.3cm}

For type IIB compactifications to AdS$_4$  the ten-dimensional supersymmetry variations  are equivalent to the following set of equations on the pure spinors $\Phi_\pm$ \cite{Grana:2004bg} 
\bea
\label{eqmapp}
&&(\d - H  \w )(e^{2A - \phi} \Phi_-)  = - 2 \mu e^{A-\phi} {\rm Re } \Phi_+,  \\
\label{eqpaapp}
&&  (\d - H \w )(e^{A -\phi } {\rm Re} \, \Phi_+) =0 \, , \\
\label{eqpbapp}
&& (\d - H \w )(e^{3 A -\phi } {\rm Im} \, \Phi_+) = -3e^{2A-\phi}{\rm Im} \, (\bar{\mu}\Phi_-) -\frac{1}{8} e^{4 A} \ast \lambda(F) \,, 
\eea
 where $\phi$ is the dilaton, $A$ the warp factor and $F$  is the sum of the RR field strength on  $Y$,  $F = F_1 + F_3 + F_5$.  $\lambda$ is the transposition on a form
 \beq
 \lambda(F_k) = (-)^{[k/2]} F_k \, .
 \eeq
It can also be shown \cite{Grana:2004bg} that for AdS$_4$ vacua supersymmetry also requires the norms of the two six-dimensional spinors to be equal
\bea
&& \| \eta^1_+ \|^2 = \| \eta^2_+\|^2 \qquad \Rightarrow \qquad |a|^2 = |b|^2 = e^A \, .  
\eea

In the following we set 
\beq
 a= \bar{b}  \qquad \,,\qquad  b = a e^{- i \theta} \, .
\eeq

Plugging the explicit form of  \eqref{Phim} and \eqref{Phip},  into the susy variations  \eqref{eqm}-\eqref{eqpb}, one can deduce  equations of definite form degree
for the forms  $z$, $\omega$ and $j$ and the fluxes. These give a set of general conditions for AdS$_4$ $\mathcal{N}=1$ susy vacua.

\subsection{No AdS$_4$ vacua for $\SU(3)$ structure}

From equation \eqref{eqm} it is immediate to see that it is not possible to have AdS$_4$ vacua with $\SU(3)$ structure.  Indeed, in this case the $\Phi_-$ only contains a three-form term, 
so that one has to zero the zero- and two-form terms in ${\rm Re } \Phi_+$, which for $\kper= 0$,  give
\bea
\label{SU3susycond}
&&  \cos \theta  = 0  \, , \nn \\
&&    \sin  \theta  (j +  \frac{i}{2} z \wedge \bar{z} ) = 0 \, .
\eea
Clearly these two equations cannot be solve at the same time.

\subsection{AdS$_4$ vacua for $\SU(2)$ structures}

Let us consider then the most general pure spinors defined in \eqref{Phimapp} and  \eqref{Phipapp}, and   first expand \eqref{eqmapp}
\beq
\label{eqmapp}
(\d - H  \w )(e^{2A - \phi} \Phi_-)  = - 2 \mu e^{A-\phi} {\rm Re } \Phi_+ \, .
\eeq
The zero-form  component gives
\beq
\mu \,  \kpar \cos \theta  = 0 \, , 
\eeq
which implies
\beq
\kpar = 0 \qquad \mbox{or} \qquad \cos \theta  = 0 \, .
\eeq
The first choice corresponds to a rigid $\SU(2)$ structure, while the second fixes the relative phase of $a$ and $b$.

\subsubsection{Rigid  $\SU(2)$ structures}

Fixing the phase of $a$ and $b$ is equivalent to fixing the orientifold projection. To allow for both O5 and O7 planes, as we need in this paper,
we have to set $\kpar = 0$ and concentrate on rigid $\SU(2)$ structure 
\beq
\kpar = 0 \qquad \kper =1 \, .
\eeq

\vspace{0.3cm}

It is convenient to define the new two-form
 \bea
\hat{\omega} =  e^{i \theta} \omega \, .
 \eea
With this redefinition, the two-, four- and six-form components of \eqref{eqmapp}  give
\bea
\label{2formapp}
&& \d (e^{3 A - \phi} z ) =  2 \mu e^{2 A - \phi}  {\rm Im} \, \hat{\omega} \,  , \\ 
\label{4formapp}
&&  \d (e^{3 A - \phi} z \w j  ) = i  e^{2 A - \phi}  H \w z  +  e^{2 A - \phi}  \mu \, z \w \bar{z} \w  {\rm Re}  \hat{\omega} \, , \\
\label{6formapp}
&&  \d (e^{3 A - \phi} z \w j \w j  ) =  2 i  e^{3 A - \phi}  H \w \hat{z} \w  j   \, .
\eea
Plugging \eqref{2formapp} in \eqref{4formapp} and recalling that ${\rm Im \omega} \w j = 0$ for an $\SU(2)$ structure we obtain
\beq
\label{4formbapp}
z \w ( \d j - i H + \mu e^{-A} \, \bar{z} {\rm Re} \,   \hat{\omega} ) =0\,. 
\eeq
It is also straightforward to show that  \eqref{6formapp} is  implied by \eqref{4formapp}. Indeed substituting   \eqref{4formapp} in \eqref{6formapp} gives
\beq
z \w j \w ( \d j - i H) = 0 \,,
\eeq
which is a consequence of  \eqref{4formbapp}.

Let us now consider the  second equation, \eqref{eqpaapp}, 
\beq
\label{eqpaapp}
 (\d - H \w )(e^{A -\phi } {\rm Re} \, \Phi_+) =0 \, .
\eeq
Expanded in forms it gives a three- and five-form equation
\bea
\label{3formaapp}
&& \d (e^{2 A - \phi}  {\rm Im} \hat{\omega} ) = 0 \, , \\
\label{5formaapp}
&&  \d( e^{2 A - \phi}  z \w \bar{z}  \w  {\rm Re} \hat{\omega} ) = 2 i  e^{2 A - \phi}   H \w  {\rm Im} \hat{\omega}\,. 
\eea

Finally we have to expand \eqref{eqpbapp}
\beq
\label{eqpbapp}
(\d - H \w )(e^{3 A -\phi } {\rm Im} \, \Phi_+) = -3e^{A-\phi}{\rm Im} \, (\bar{\mu}\Phi_-)-\frac{1}{8} e^{4 A} \ast \lambda(F) \,. 
\eeq
This gives
\bea
\label{1formbapp}
&& \ast F_5 =  3 e^{-A - \phi} \, {\rm Im } (\bar{\mu} z) \, , \\
\label{3formbapp}
&& \ast F_3 = -e ^{ -4 A} \, \d (e^{4 A - \phi} {\rm Re} \hat{\omega} ) +3 e^{-A - \phi} {\rm Re} (\bar{\mu} z) \w j  \, , \\
\label{5formbapp}
&&  \ast F_1 =  -i \, \d (2 A - \phi)  z \w \bar{z}  \w {\rm Im} \hat{\omega}  -   e^{-\phi} H \w {\rm Re} \hat{\omega} \nn \\
& &  \qquad  \quad +
\frac{1}{2} e^{-A - \phi} {\rm Im} (\bar{\mu} z ) \w j \w j \,,
\eea
where in the last equation we used \eqref{3formaapp}. In summary the non trivial susy conditions are
\bea
\label{Phimexpf1}
&& \d (e^{3 A - \phi} z ) =  2 \mu e^{2 A - \phi}  {\rm Im} \, \hat{\omega} \,  , \\ 
\label{Phimexpf2}
&& z \w ( \d j - i H + \mu e^{-A} \, z \w  {\rm Re} \,   \hat{\omega} ) =0  \,,\\
\label{Phipaexpf1}
&& \d (e^{2 A - \phi}  {\rm Im} \hat{\omega} ) = 0 \, , \\
\label{Phipaexpf2}
&&  \d( e^{2 A - \phi}  z \w \bar{z}  \w  {\rm Re} \hat{\omega} ) = 2 i  e^{2 A - \phi}   H \w  {\rm Im} \hat{\omega} \,,
\eea
plus  equations \eqref{1formbapp}-\eqref{5formbapp} for the fluxes.

\vspace{0.3cm}

To make contact with previous literature,  we can express the equations above using the    $\SU(2)$ intrinsic torsions \eqref{SU2torsions}. The idea is to decompose all the
objects in the equations in representations of $SU(2)$ and then obtain a set of conditions for the fields in the various representations.
To decompose the exterior derivatives we use the  torsion classes  defined in \eqref{SU2torsions}, while for the fluxes we have
\begin{align}
 H= &h_1 z \wedge \hat{\omega}+h_2\bar{z} \wedge \hat{\omega}
+h_3 z  \wedge j+ z \wedge \bar{z} \wedge h_1^{(2)}+h_{2}^{(2)}\wedge j + z \wedge  h^{(3)}+\mathrm{c.c.}\,,\\
F_1=&f_1 z+f_1^{(2)}+\mathrm{c.c.}\,,\\
F_3=&f_2 z \ \wedge \hat{\omega}+f_3\bar{z} \wedge  \hat{\omega}+f_4 z \wedge j+ z \wedge \bar{z} \wedge f_2^{(2)}+f_3^{(2)} \wedge j+ z \wedge f^{(3)}+\mathrm{c.c.}\,,\\
F_5=&f_5 z \wedge j \wedge j + z \wedge \bar{z}  \wedge j \wedge f_4^{(2)}+\mathrm{c.c.}\,,
\end{align}
where  $h_i$ and $f_i$ are complex scalars in the singlet representation of $\SU(2)$,  $h_i^{(2)}$ and $f_i^{(2)}$ are holomorphic vectors in  the ${\bf 2}$ and $h^{(3)}$ and
$f^{(3)}$ are complex two forms in the triplet representation, which are (1,1) and primitive with respect to $j$.

For completeness we also give the decomposition of Hodge dual fluxes
\begin{align}
 \ast H=&-ih_1 z \wedge \hat{\omega}+i h_2\bar{z} \wedge \hat{\omega}-i h_3 z \wedge j-i z \wedge  \ast_4h^{(3)}+2i\ast_4h_1^{(2)} \nn  \\
 & -\frac{i}{2} z \wedge \bar{z}(h_2^{(2)}\llcorner j)+\mathrm{c.c.}\,,\\
\ast F_1=&-\frac{i}{2}f_1 z \wedge j \wedge j - \frac{i}{2} z \wedge \bar{z} \wedge \ast_4f_1^{(2)}+\mathrm{c.c.}\,,\\
\ast F_3=&-if_2 z \wedge \hat{\omega}+if_3\bar{z} \wedge \hat{\omega}-if_4 z \wedge j-i z \wedge \ast_4 f^{(3)}+2i\ast_4f_2^{(2)}\nn \\
& -\frac{i}{2} z \wedge \bar{z} \wedge (f_3^{(2)}\llcorner j)+\mathrm{c.c.}\,,\\
\ast F_5=&-2i f_5 z+2i f_4^{(2)}\llcorner j+\mathrm{c.c.}\,,
\end{align}
where we used the fact that a product structure allows to split  $\ast_6=\ast_2\ast_4$ and $\ast_4(v\wedge\xi)=(-1)^{\mathrm{deg}\xi}v\llcorner\ast_4\xi$.

We can now look at the SUSY variations. Let us first consider \eqref{Phimexpf1} - \eqref{Phipaexpf2}.  We find that the singlets in the  torsions must satisfy 
\beq
\label{singlets}
\begin{array}{lll}
S_2= 0\,, &  \qquad  S_1 = - S_4 = -i \mu e^{-A} \,, \\
S_3=\frac{1}{2}\partial_{\bar{z}}(3A-\phi)\,,   & \qquad   S_5 = \bar{S}_6 = i \bar{h}_1 -\frac{1}{2} e^{-A}\mu \,, \\
& \qquad S_7=\bar{S_8} = - \frac{1}{2} \partial_{z}(2A-\phi)   \, .  & 
\end{array}
\eeq
Similarly, there are conditions on  the vectors  
\beq
\label{vectors}
\begin{array}{ll}
V_3=V_4= V_6 = 0 \,,   & \quad   V_7  = i  (\bar{\partial}_4 A  + \bar{h}^{(2)}_1 )  \llcorner  \omega \,, \\  
V_5=i h_2^{(2)} \,,  &  \quad V_8 = i     [ \bar{\partial}_4 (3 A -\phi)  + \bar{h}^{(2)}_1 ]  \llcorner  \omega \,, \\
 V_1=V_2=\partial_4 (3A-\phi)\,,   & \qquad  
\end{array}
\eeq
and the two-forms 
\beq
\label{2forms}
T_1 = 0 \,,  \qquad T_2 = - i h^{(3)} \,,\qquad T_3 = \bar{T}_4  \, ,
\eeq
and the NS flux singlets
\beq
h_1 =\bar{h}_2\,, \qquad h_3 =  -  \frac{i}{2} \partial_z (2 A - \phi) \, .
\eeq

Finally the equations \eqref{1formbapp} - \eqref{5formbapp} for the RR fluxes give
\beq
\label{fluxcon}
\begin{array}{lll}
 f_1 =-  e^{- \phi} ( 4 i h_1 - \frac{1}{2} \bar{\mu} e^{-A} ) \,, & \qquad f^{(2)}_1 = i e^{- \phi}  \omega \llcorner [\bar{\partial}_4 (2 A - \phi) + \bar{ h}^{(2)}_1 ]  \,, \\
 f_2 = \bar{f}_3 = - \frac{i}{2} e^{- \phi} \partial_z A\,,  & \qquad   f^{(2)}_2 =  \frac{i}{2} e^{- \phi} \omega \llcorner  [\bar{\partial}_4 (4 A - \phi) - \bar{ h}^{(2)}_1 ]  \,,\\
f_4 =  \frac{1}{2}  e^{- \phi} ( 4 h_1 +  i  \bar{\mu} e^{-A} )   & \qquad   f^{(2)}_3 = f^{(2)}_4 = 0 \,, \\
f_5 =  \frac{3}{4} e^{- A - \phi} \bar{\mu}  & \qquad f^{(3)} =  i  e^{-\phi} T_3  \,.
\end{array}
\eeq

The conditions coming from the Bianchi identities are too involved to give for a generic $\SU(2)$ structure. They become more amenable in presence of O-planes,
since the orientifold projections considerably restrict the allowed torsion classes. We discuss the case of parallelisable  manifolds in the main text.

\section{Type IIA $AdS_4$ solutions and separation of scales} 
\label{IIAapp}

The best known examples of AdS$_4$ compactifications and, among them, AdS$_4$ vacua with separation of scales are 
in type IIA. In this Appendix we first remind the general conditions for $\mathcal{N}=1$ AdS$_4$ vacua in type type IIA  \cite{Lust:2004ig} 
and then discuss separation of scales for a specific example.

\subsection{Lust--Tsimpis $\SU(3)$-structures}
\label{LT}

The general conditions for AdS$_4$ with $\mathcal{N}=1$ supersymmetry on $\SU(3)$ structure manifolds given in  \cite{Lust:2004ig}  is
easily derived using the pure spinor equations   \cite{Grana:2006kf}.

In type IIA the supersymmetry variations take the same form as in \eqref{eqmapp}-\eqref{eqpbapp} with $\Phi_+$ and $\Phi_-$ exchanged
\bea
\label{eqpIIA}
&&(\d - H  \w )(e^{2A - \phi} \Phi_+)  = - 2 \mu e^{A-\phi} {\rm Re } \Phi_-,  \\
\label{eqmaIIA}
&&  (\d - H \w )(e^{A -\phi } {\rm Re} \, \Phi_-) =0 \, , \\
\label{eqmbIIA}
&& (\d - H \w )(e^{3 A -\phi } {\rm Im} \, \Phi_-) = -3e^{2A-\phi}{\rm Im} \, (\bar{\mu}\Phi_+) -\frac{1}{8} e^{4 A} \ast \lambda(F) \,.
\eea

For an $\SU(3)$ structure, the pure spinors have the form \eqref{SU3pstext}
\be
\Phi_- = -\frac{iab}{8}\Omega  \qquad \Phi_+ = \frac{a\bar{b}}{8}e^{-iJ}
\ee
where we set  $\e^{A/2}=|a|=|b|=1$.  It is convenient to define 
\be
\bar{a}b\mu\equiv \hat{\mu}\equiv m + i \tilde{m}\,, \qquad \hat{\Omega}=-i ab\Omega\,,
\ee 
where number $m$ is not to be confused with the Romans mass, although it will turn out to be proportional to it. Then
real and imaginary parts of the  pure spinors can be written as
\begin{align}
& 8 \,  {\rm Re} (\bar{\mu}\Phi_+) = m - \tilde{m}J - m\tfrac{1}{2}J^2 + \tilde{m}\tfrac{1}{3!}J^3\,,\\	
& 8 \,  {\rm Im}(\bar{\mu}\Phi_+) = -\tilde{m} -  m J + \tilde{m}\tfrac{1}{2}J^2 + m\tfrac{1}{3!}J^3\,,\\
& 8 \,  {\rm Re}(\Phi_-)= \hat{\Omega}_R\,,\\
& 8 \, {\rm Im}(\Phi_-) = \hat{\Omega}_I\,.
\end{align}

The supersymmetry variations \eqref{eqpIIA} and \eqref{eqmaIIA}  imply
\be
 H=2m \hat{\Omega}_R\,,\qquad \d J= 2\tilde{m} \hat{\Omega}_R\,.
\ee
This leads to the following restrictions on the torsions
\begin{align}
& W_1=-\frac{4}{3}i\tilde{m}\,,\qquad W_4=0\,,\qquad W_3=0\,,\nonumber\\
&{\rm Re}(W_2)=0\,,\qquad {\rm Re}( \bar{W}_5 \hat{\Omega})=0\,.
\end{align}
The last pure spinor equation, \eqref{eqmbIIA} defines the RR field strengths
\begin{align}
&\e^{\phi} F_0 = 5 m\,,\\
&\e^{\phi} F_2 =\frac{\tilde{m}}{3}J  +i W_2\,,\\
&\e^{\phi} F_4 =\tfrac{3}{2}mJ\w J\,,\\
&\e^{\phi} F_6 =3\tilde{m} \, {\rm vol}_6\,.
\end{align}
The Bianchi identity for $F_4$ implies that $\hat{\Omega}_R\wedge \ast {\rm Im}(\bar{W}_5 \hat{\Omega}) =0$,
where we made use of $\ast W_2 = J \wedge W_2$. Together with ${\rm Re}( \bar{W}_5 \hat{\Omega})=0$ we find $W_5=0$. 
The $F_2$ Bianchi identity $\d F_2 = F_0 H +j$ gives an equation for the source form
\be
\e^{\phi}j_{\rm  O6} = i\d W_2 + (\frac{2}{3}\tilde{m}^2 -10 m^2) \hat{\Omega}_R\,.
\ee
Source-less solutions require 
\be\label{sourcefree}
i\d W_2= c\Omega_R\,,\qquad c  =10 m^2 -   \frac{2}{3}\tilde{m}^2\,.
\ee
One can easily prove that if $i\d W_2= c\Omega_R$ we must have that\footnote{This is done by considering the identity $\d(\Omega_I\wedge W_2)=0$ and then using the various $\SU(3)$-structure identities on this.}
\be
c=-\tfrac{1}{8}(W_2^I)_{ab}(W_2^I)^{ab}\,.
\ee
Explicit manifolds for which these conditions can be fulfilled have been constructed \cite{Koerber:2008rx}.

\vspace{0.3cm}

For simplicity we restrict to the class of the solutions that obey $i\d W_2=c\Omega_R$. That class has the nice property that it connects to the source-free case for special value of $c$  given in (\ref{sourcefree}). The source form $j$ that appears in the Bianchi identity is then proportional to $\Omega_R$ 
\be
e^{\phi}j_{\rm O6} = Q\Omega_R
\,,\ee
where $Q$ relates to the O6 or D6  charge. 
We now follow the argument of \cite{Tsimpis:2012tu}. If we use that $R_6=\tfrac{15}{2}|W_1|^2 -\tfrac{1}{2}|W_2|^2$, we can derive that
\be
r= \frac{6\tilde{m}^2 + 10m^2}{m^2+\tilde{m}^2} + \frac{Q}{m^2 +\tilde{m}^2}\,.
\ee
The first term is clearly bounded, so scale separation then indeed requires non-zero orientifold charge and small cosmological constant.

\subsection{The rectangular torus solution}

A well known example where the conditions for a classical  $AdS_4$ vacuum with scale separation are 
satisfied is provided by  the model of   \cite{DeWolfe:2005uu}:  a compactification
on an orbifold of $T^6$ with  non-zero $F_0, F_4, H$-fluxes and O6 sources. 
While the $H$ and $F_0$ flux are constrained by the tadpole condition to be of order one,  the $F_4$ flux is an unbounded flux quantum.  
All moduli are stabilised  and, by scaling up the $F_4$ flux quanta,  the conditions
for scale separation can be met. Since $F_4$   is  unbounded,  there is actually an infinite number  of solutions.  Since the internal space has no curvature  only criterium (\ref{scalesep1}), but not (\ref{scalesep2}) can  be used.

\vspace{0.3cm}
 
In order to simplify things we describe the solution on the rectangular torus, by which we mean that only the dependence on the volume modulus is shown since all other moduli are fixed to their miniumum for which the torus is straight. The solution is given by
\begin{align}
& J = \rho^2 [e^{12}+e^{34}+e^{56}] \,,\\
& \Omega = \rho^3 (ie^1+e^2)\wedge (ie^3+e^4)\wedge(ie^5+e^6)\,,\\
&  {\rm vol}_6 = -\rho^6 e^{123456}\,,
\end{align}
where $e^i=\d x^i$ with the $x^i$ the local Cartesian coordinates on a torus. 
All torsion classes are zero. The 10D solution is given by
\begin{align}
& e^{\phi}F_0 = 5m \,,\\
& e^{\phi}F_4 = \frac{3}{2}m J\wedge J \,,\\
& H = 2 m \Omega_R \,,\\
& e^{-\phi}j_{O6} = -10m^2 \Omega_R
\end{align}
The cosmological constant is given by
$ V=  -6m^2M_p^2$. 

Now we discuss the scalings and the limit to weak coupling, large volume, small cc, scale separation and we take into account flux and charge quantisation. 
We introduce a parameter $\lambda$ which is supposed to go to infinity
\be
\lambda \rightarrow \infty\,.
\ee
Small cc can be achieved by making $m$ scale as follows
\be
m \rightarrow \lambda^{-1}\,.
\ee
The quantisation of Romans mass $F_0 = n_0$ and the $F_2$ tadpole condition 
\be
n_0 h = \text{number of O-planes}\,,
\ee
imply that both Romans mass quantum and the $H$ flux quantum $h$ are bounded to be order one. We then find, from the expression of $F_0$ that
\be
e^{\phi} \rightarrow \lambda^{-1}\,.
\ee
This ensures weak coupling. The boundedness of the $h$ flux quantum together with the explicit expression for $H$ flux implies that
\be
\rho \sim \lambda^{1/3}\,,
\ee
This brings us to large volume.  Flux quantisation for $F_4$ 
\be
F_4 = n_4 (\text{wedges of e's})\,,
\ee
implies that $n_4$ will become a large flux quantum 
\be
n_4 \sim \lambda^{2/3}\,.
\ee
Note that the  $F_4$  flux quanta  are  not bounded by a tadpole. The scalings of the fluxes are nicely consistent with keeping the number of O6 planes to be order one since
\be
j_{O6}\sim e^{-\phi} m^2 \rho^3 (e^{246} - e^{136} - e^{235} + e^{145} ) \,,
\ee
where
\be
 e^{-\phi} m^2 \rho^3 \rightarrow \lambda^0  =\mathcal{O}(1)\,.
\ee
Taking
\be
L^2_{AdS} = m^{-2}\rightarrow \lambda^2 \,, \qquad L^2_{KK} = \gamma^2\rightarrow \lambda^{2/3}\,,
\ee
we find scale separation 
\be
L^2_{AdS}/L^2_{KK}\rightarrow \lambda^{4/3}\rightarrow \infty\,.
\ee

\section{Computation of the 6D Ricci scalar}
\label{riccisapp}

In this section we give the explicit computation of the Ricci scalar in terms of $\SU(2)$ torsion classes for the manifolds we consider in the main text. 
In order to do that,  we use the fact that, given an $\SU(2)$ structure, it is always possible to embed it in an $\SU(3)$ structure. In this case we choose
\be
 J=j - z_R\wedge z_I\qquad \Omega=\omega\wedge z\,.
\ee
Then we can express the $\SU(3)$ torsion classes \eqref{SU3torsions} in terms of  the $\SU(2)$ ones \eqref{SU2torsions}. This can be in general be very cumbersome, but  it simplyfies a lot   for the $\SU(2)$ structures compatible with the O5/O7 projections, since many of the torsion  classes are projected out.  Comparing \eqref{SU3torsions}  and (\ref{SU2torsionspar}), one can show that 
\begin{align}
\label{SU3toSU2}
 W_1=&- \frac{2}{3}(S_1 + 2 i S_5 )\,,\\
W_2=&- \frac{4}{3} ( S_1 - i S_5) ( j - 2 z_R\wedge z_I) +2i\bar{T}_3\,, \\
W_3=&-2\Im(\bar{h}_1 z\wedge \omega)+2\Im(z\wedge h^{(3)})\,,\\
W_4 = &\, 0 = W_5\,.
\end{align}
The expression of the Ricci scalar in terms  of the $\SU(3)$ torsion classes is known \cite{Ali:2006gd, bedulli-2007-4}. When both $W_4$ and $W_5$ vanish the Ricci scalar is 
\beq
 R_6=\frac{15}{2}|W_1|^2-\frac{1}{2}|W_2|^2-\frac{1}{2}|W_3|^2 \, .
 \eeq
Then using \eqref{SU3toSU2} we obtain
\bea
R_6&=&-4(-3|e^{-A}\mu|^2+8\Im(e^{-A}\mu h_1)+|h^{(3)}|^2+|T_3|^2)\nonumber\,,\\
&=&-4(|S_1|^2 +  4 |S_5|^2 + \Im (S_5 \bar{S}_1) + |T_2|^2 + |T_3|^2 )\,.
\eea

Also note that in sourceless solution, one has (using (\ref{RRBIs})) 
\beq 
 R_6=10|e^{-A}\mu|^2+32|h_1|^2>0 \, .
\eeq
We recover the fact that  the manifold must have a positive internal curvature in order to have sourceless SUSY solution.

\bibliographystyle{utphysmodb}
\bibliography{SUSYAdS}
\end{document}